\pdfoutput=1 
\documentclass[journal,twoside,web]{ieeecolor}
\usepackage{tmi}
\usepackage{cite}
\usepackage{amsmath,amssymb,amsfonts}
\usepackage{algorithmic}
\usepackage{graphicx}
\usepackage{textcomp}
\usepackage{multirow}
\usepackage{hyperref}
\usepackage{diagbox}
\usepackage{booktabs}
\usepackage{caption}
\usepackage{subcaption}
\usepackage{pifont}

\newcommand{\revise}{}

\def\etal{\MakeLowercase{\textit{et al.}}}
\def\ie{\MakeLowercase{\textit{i.e.}}}
\def\BibTeX{{\rm B\kern-.05em{\sc i\kern-.025em b}\kern-.08em
    T\kern-.1667em\lower.7ex\hbox{E}\kern-.125emX}}
\begin{document}
\title{One Model to Synthesize Them All: \\ Multi-contrast Multi-scale Transformer for Missing Data Imputation}
\author{Jiang Liu$^{*}$, \IEEEmembership{Graduate Student Member, IEEE}, Srivathsa Pasumarthi$^{*}$, Ben Duffy, Enhao Gong, Keshav Datta, and Greg Zaharchuk.
\thanks{ $^{*}$The first two authors are equal contributors and co-first authors. Manuscript initially submitted on 4/27/2022, resubmitted with major revisions on 02/16/2023, and accepted on 03/21/2023. This work is supported by NIH grant R44EB027560. We thank Dr. Rama Chellappa (Life Fellow, IEEE) for his helpful comments. We also thank Dr. Thomas Campbell Arnold and Dr. Ryan Chamberlain for their valuable inputs. } 
\thanks{Jiang Liu (jiangliu@jhu.edu) is with the Department of Electrical and Computer Engineering, Johns Hopkins University, Baltimore, MD, USA. This work was done as part of his internship with Subtle Medical Inc.}
\thanks{Srivathsa Pasumarthi, Dr. Ben Duffy, Dr. Keshav Datta and Dr. Enhao Gong (srivathsa, ben, keshav, enhao@subtlemedical.com) are with Subtle Medical Inc., Menlo Park, CA, USA.}
\thanks{Dr. Greg Zaharchuk (gregz@stanford.edu) is with the Department of Radiology, Stanford University, CA, USA. }
\thanks{A part of this work was presented as a Digital Poster in the 31st Annual Conference of the International Society of Magnetic Resonance in Medicine (ISMRM), May 7-12, 2022, London, UK.}
}

\maketitle

\begin{abstract}
Multi-contrast magnetic resonance imaging (MRI) is widely used in clinical practice as each contrast provides complementary information. However, the availability of each imaging contrast may vary amongst patients, which poses challenges to radiologists and automated image analysis algorithms. A general approach for tackling this problem is missing data imputation, which aims to synthesize the missing contrasts from existing ones. While several convolutional neural networks (CNN) based algorithms have been proposed, they suffer from the fundamental limitations of CNN models, such as the requirement for fixed numbers of input and output channels, the inability to capture long-range dependencies, and the lack of interpretability. In this work, we formulate missing data imputation as a sequence-to-sequence learning problem and propose a multi-contrast multi-scale Transformer (MMT), which can take \textit{any} subset of input contrasts and synthesize those that are missing. MMT consists of a multi-scale Transformer encoder that builds hierarchical representations of inputs combined with a multi-scale Transformer decoder that generates the outputs in a coarse-to-fine fashion. The proposed multi-contrast Swin Transformer blocks can efficiently capture intra- and inter-contrast dependencies for accurate image synthesis. Moreover, MMT is inherently interpretable as it allows us to understand the importance of each input contrast in different regions by analyzing the in-built attention maps of Transformer blocks in the decoder. Extensive experiments on two large-scale multi-contrast MRI datasets demonstrate that MMT outperforms the state-of-the-art methods quantitatively and qualitatively. 

\end{abstract}

\section{Introduction}
\label{sec:introduction}
\IEEEPARstart{M}{agnetic} resonance imaging (MRI) can visualize different soft tissue characteristics by varying the sequence parameters, such as the echo time and repetition time. Through such variations, the same anatomical region can be visualized under different contrast conditions, and the collection of such images of a single subject is known as multi-contrast MRI. Multi-contrast MRI provides complementary information about the underlying structure as each contrast highlights different anatomy or pathology. It is widely used in different clinical applications \cite{mpmri_brain, mpmri_prostate} across organs in the human body as well as for downstream quantification tasks such as tumor segmentation \cite{tumorseg_overview}, vertebral disc segmentation \cite{ivd_seg}, and parametric mapping for myocardial diseases \cite{myocard}. However, the availability of each contrast may vary amongst patients in practice, especially in large-scale multi-institution studies, due to issues including prohibitive scanning time, image corruption (for example, due to motion), and different acquisition protocols which pose challenges for both radiologists and automated image analysis algorithms.

\begin{figure*}[htbp]
  \centering
  \includegraphics[width=1.0\textwidth]{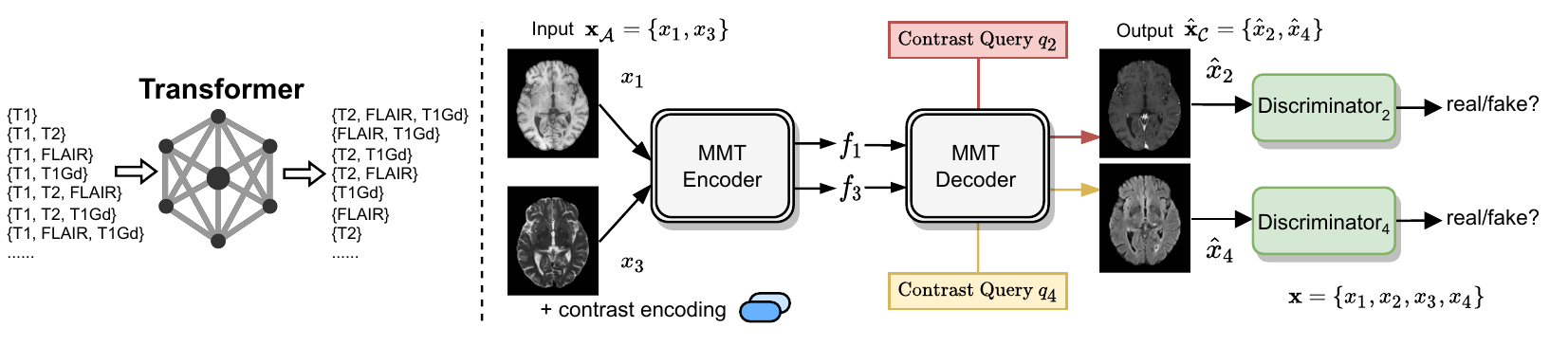}
  \caption{Left: We formulate missing data imputation as a sequence-to-sequence prediction problem and solve it with a Transformer network. Right: Overview of the proposed multi-contrast multi-scale Transformer (MMT) network. MMT consists of a \revise{\textit{multi-scale}} Transformer encoder \revise{(Fig.~\ref{fig:encoder})} that builds \revise{\textit{hierarchical}} representations of inputs, and a \revise{\textit{multi-scale}} Transformer decoder \revise{(Fig.~\ref{fig:decoder})} that generates the outputs in a \revise{\textit{coarse-to-fine}} fashion. Given a learned target contrast query, MMT reasons about the input-target contrast relationship and considers the local and global image context to generate the final synthetic image. } 
  \label{fig:overview}
\end{figure*}


\looseness -1 A general approach for tackling the missing contrast problem is missing data imputation, which aims to synthesize the missing contrasts from existing ones to avoid the cost of rescanning or modifying existing pipelines for handling missing contrasts. To fully utilize the available information for accurate synthesis, a missing data imputation algorithm should take all available contrast(s) as input to extract the complementary information and output the missing contrast(s), which can be many-to-one, one-to-many and many-to-many synthesis depending on the number of available and unobtained/corrupted contrasts.

\revise{Efficiently solving the missing data imputation problem is crucial in a real-world clinical setting and can have many impactful applications. The ability of such models to synthesize an entirely new sequence from an existing sequence or sequences can avoid rescans and reduce the number of scans with quality and motion issues. Apart from overall motion mitigation, missing data imputation models can synthesize targeted sequences like MRA-TOF from T1/T2 brain images \cite{mrasynth}, T1Gd from pre-contrast \cite{reb_gad} and STIR from T1/T2 spine images \cite{spinesynth}. These techniques reduce the scan time, improve the overall patient experience, and reduce operational costs.}

\looseness -1 While several convolutional neural networks (CNN)
based missing data imputation algorithms~\cite{milr, mmgan, remic} have been proposed, they suffer from the fundamental limitations of CNN models. First of all, CNNs require fixed input and output channels. In the worst case, we need to train $(2^{P}-2)$ models, one for each possible input-output scenario, where $P$ is the number of contrasts. Although several strategies have been proposed to circumvent this issue, including feature map fusion and pre-imputation, they are not ideal. Feature map fusion~\cite{milr} fuses the feature maps of input contrasts by a \texttt{Max($\cdot$)} function, such that the input to the decoder network always has the same number of channels regardless of the number of input contrasts. However, the input contrasts are encoded separately, and the predefined \texttt{Max($\cdot$)} function does not necessarily capture the complementary information within each contrast. Pre-imputation~\cite{mmgan, remic} pre-imputes missing contrasts with zeros such that the input and output of synthesis networks always have $P$ channels. As a result, the network may consider each input contrast independently instead of exploring complementary information, as any input channel can be zero. Moreover, CNNs are not good at capturing the long-range dependencies within the input images since they are based on local filtering, while spatially distant voxels in medical images can have strong correlations and provide helpful information for synthesis. Another concern of typical CNNs is their lack of interpretability; \ie, there is no explanation about why they produce a particular image and where the information comes from, which is vital for building trustworthy medical imaging applications \cite{ai_expln_trust}. Although several techniques have been proposed for posthoc interpretability analysis for CNNs~\cite{8237336, zeiler2014visualizing, yosinski2015understanding}, they do not explain the reasoning process of how a network actually makes its decisions~\cite{chen2019looks}.   

\looseness -1 In this paper, we formulate missing data imputation
as a \textit{sequence-to-sequence} prediction problem, \revise{\ie, mapping a sequence of input (available) contrasts of \textit{variable} lengths and combinations to the sequence of output (missing) contrasts. This formulation is flexible and captures the random nature of the missing data imputation problem that any subset of the contrasts can be missing. This is in contrast with the existing \textit{image-to-image} translation framework \cite{milr, mmgan, remic}, where a model transforms a \textit{fixed} set of input images to the output images. Under this formulation, the missing data imputation problem can be well solved with Transformer models since they can naturally handle input and output sequences of arbitrary length thanks to the attention mechanism to deal with exponentially many input-output scenarios with only one model. }

\revise{To this end, we design a novel Multi-contrast Multi-scale Transformer (MMT). MMT is based on the proposed Multi-contrast Shifted Window based Attention (M-Swin), where attention computation is performed within local cross-contrast windows. M-Swin generalizes the shifted window approach in Swin Transformer~\cite{swin} and allows us to capture intra- and inter-contrast dependencies for accurate synthesis efficiently.} The overview of the proposed MMT network is shown in Fig.~\ref{fig:overview}. \revise{ MMT consists of a multi-scale Transformer encoder that builds hierarchical representations of inputs and a multi-scale Transformer decoder that generates the outputs in a coarse-to-fine fashion. } Given a learned target contrast query, MMT reasons about the input-target contrast relationship and considers the local and global image context to generate the final synthetic image. The attention scores inside the MMT decoder indicate the amount of information coming from different input contrasts and regions for synthesizing every pixel in the output image, which makes MMT inherently interpretable. We perform extensive experiments on the IXI~\cite{ixi} and BraTS 2021~\cite{baid2021rsna} datasets and demonstrate that MMT outperforms the state-of-the-art methods both quantitatively and qualitatively. We further demonstrate one real-world application of MMT by using imputed images for tumor segmentation \revise{and perform radiomics feature analysis to show the diagnostic value of the imputed images}.

The summary of our contributions is as follows:
\begin{enumerate}
    \item We formulate missing data imputation as a sequence-to-sequence prediction problem and propose a novel Transformer-based framework that can take any subset of contrasts as input and generate the missing ones.
    \item We design a novel multi-contrast multi-scale Transformer (MMT),  which builds hierarchical representations of multi-contrast inputs and generates the outputs in a coarse-to-fine fashion with inherent interpretability.
    \item We propose Multi-contrast Shifted Window based Attention (M-Swin), which efficiently captures intra- and inter-contrast dependencies for accurate synthesis.
    \item Our extensive experiments on two large-scale multi-contrast MRI datasets demonstrate the superiority of the proposed method quantitatively and qualitatively. 
\end{enumerate}

\section{Related Work}
\subsection{MRI image synthesis}
\label{sec:image_syn}
Synthesizing new contrast information from available images has been an active area of research in the MRI domain. Traditional methods includes patch regression~\cite{r1}, atlas based approaches~\cite{r3, r4}, and MR physics based approaches~\cite{r5, r6}. 
In the recent years, many CNN-based approaches have demonstrated state-of-the-art performance for MR contrast synthesis~\cite{r8, r9, r10, r11, pGAN,  r13, r14, r15,  r18, mustgan, collagan, diamondgan, ipmi, 10.1007/978-3-031-17027-0_6}. Generative adversarial networks were introduced to enable the synthesis of realistic images across various domains~\cite{pGAN, r15, r18}. Dar \etal~\cite{pGAN} proposed pGAN based on conditional GAN~\cite{r13} and cGAN based on CycleGAN~\cite{r14}. \cite{r15} used a style transfer GAN for contrast-aware synthesis. \cite{r18} proposed a 3D self-attention conditional GAN with spectral normalization for synthesizing mean diffusivity maps from multi-modal MRI images.
While most of the literature focused on \textit{one-to-one} synthesis, several studies~\cite{mustgan, collagan, diamondgan, hinet} considered the \textit{many-to-one} synthesis problem, where the algorithm takes multiple contrasts as input and generates one missing contrast. mustGAN~\cite{mustgan} fused the unique features in multiple one-to-one streams and shared features in a many-to-one stream. CollaGAN~\cite{collagan} can generate any target contrast with the other contrasts as input using a single generator by adding a mask vector indicating the desired output. DiamondGAN~\cite{diamondgan} learns the mapping between any subsets of multiple input contrasts to a target contrast \revise{with a multi-modal cycle-consistency loss function}. \revise{Hi-Net~\cite{hinet} proposed a hybrid fusion strategy to fuse the representations of various input modalities for image synthesis. Missing data imputation is the most general MRI image synthesis task, which covers \textit{one-to-one}, \textit{one-to-many}, \textit{many-to-one}, and \textit{many-to-many} synthesis depending on the number of available inputs and missing outputs, and we provide an overview of current works in Sec. \ref{sec:missing_data}.}

\subsection{Transformer}
Transformer networks were first introduced in~\cite{NIPS2017_3f5ee243} for natural language processing. A Transformer consists of two parts: an encoder that maps an input sequence of symbol representations to a sequence of continuous representations and a decoder that generates an output sequence of symbols one element at a time, given the encoder outputs.  
Transformer-based models have become the state-of-the-art methods for many computer vision tasks~\cite{vit, detr, swin, jiang2021transgan, zheng2021rethinking, chen2021pre} due to their ability to capture long-range dependencies within the input. Transformers have also been employed in various medical imaging applications, such as image segmentation~\cite{valanarasu2021medical}, image registration~\cite{chen2021vit}, and image synthesis~\cite{r26}. In the domain of MRI, Zhang \etal~\cite{r25} developed PTNet for synthesizing infant MRI images. Feng \etal~proposed T$^2$Net~\cite{feng2021task} for joint MRI reconstruction and super-resolution. \cite{korkmaz2022unsupervised} proposed SLATER for zero-shot MRI reconstruction. Dalmaz \etal \cite{resvit} proposed ResViT for missing MRI contrast synthesis.

\subsection{Missing data imputation}
\label{sec:missing_data}
Missing data imputation is the problem of estimating the value of missing data based on other available information. It is a more challenging image synthesis problem, requiring an algorithm to handle all possible missing data scenarios. The models described in Sec.~\ref{sec:image_syn} cover only a subset of different input-output configurations. To efficiently account for all scenarios with just one model, MILR~\cite{milr} uses a feature map fusion strategy, where feature maps of input contrasts are encoded separably and then fused by a \texttt{Max($\cdot$)} function for decoding target contrasts; MMGAN~\cite{mmgan} and ReMIC~\cite{remic} pre-impute missing contrasts with zeros and then feed them into a generator with existing contrasts for synthesis. Though promising, these algorithms suffer from the aforementioned fundamental limitations of CNNs, \revise{such as the requirement of a fixed number of input and output channels, the inability to capture long-range dependencies, and the lack of interpretability. ResViT~\cite{resvit} is the first missing data imputation algorithm with Transformer. However, the ResViT framework is largely based on CNN networks except for the central bottleneck, which utilizes aggregated residual Transformer blocks that combine convolutional and Transformer modules. It still needs to pre-impute missing contrasts similar to ~\cite{mmgan, remic} and lacks interpretability.}

\section{Methods}

\subsection{Overview}
The overall MMT architecture is shown in Fig.~\ref{fig:overview}. Consider MRI images of $P$ contrasts $\mathbf{x}=\{x_1, x_2, \cdots, x_P\}$. Given a sequence of arbitrary $M$ $\left(1\leq M\leq P-1\right)$ input contrasts $\mathbf{x}_{\mathcal{A}}=\{x_{a_i}\}_{i=1}^{M}$, the goal of MMT is to synthesize the remaining $N$ contrasts $\mathbf{x}_{\mathcal{C}}=\mathbf{x}\setminus\mathbf{x}_{\mathcal{A}}=\{x_{c_i}\}_{i=1}^{N}$, where $N=P-M$, and $\mathcal{A}=\{a_i\}_{i=1}^M$, $\mathcal{C}=\{c_i\}_{i=1}^{N}$ are the indices of available contrasts and missing contrasts respectively.

The MMT encoder $Enc$ first maps the input sequence $\mathbf{x}_\mathcal{A}$ to a sequence of multi-scale feature representations $\mathbf{f}_\mathcal{A}=\{f_{a_i}\}_{i=1}^{M}$, where $f_{a_i}=Enc(x_{a_i})=[f_{a_i}^{(1)}, f_{a_i}^{(2)}, \cdots, f_{a_i}^{(S)}]$ and $f_{a_i}^{(s)}$ is the feature of input contrast $x_{a_i}$ at scale $s$. 

Given $\mathbf{f}_\mathcal{A}$ and the contrast queries $\mathbf{q}_\mathcal{C}=\{q_{c_i}\}_{i=1}^{N}$ of the target contrasts, the MMT decoder $Dec$ then reasons about the input-target contrast relationship and synthesizes the output sequence $\hat{\mathbf{x}}_{\mathcal{C}}=\{\hat{x}_{c_i}\}_{i=1}^{N}$ one element at a time:
\begin{equation}
    \hat{x}_{c_i}=Dec(\mathbf{f}_\mathcal{A}; q_{c_i}),~c_i\in \mathcal{C}.
\end{equation}

The detailed architectures of the MMT encoder and decoder are presented in Sec.~\ref{sec:encoder} and Sec.~\ref{sec:decoder}. We use CNN-based discriminators to guide the training of MMT for improving image quality, which is detailed in Sec.~\ref{sec:disc}.

\revise{
MMT has the following desired properties for missing data imputation:
\begin{enumerate}
    \item \textbf{Flexibility}: MMT can take any arbitrary subset of contrasts as input and generate the missing ones.
    \item \textbf{Multi-contrast processing}: MMT can extract complementary information from multi-contrast input images, when applicable, for accurate synthesis. 
    \item \textbf{Multi-scale processing}: MMT generates hierarchical representations at multiple scales, which is shown to be beneficial in the medical imaging domain~\cite{ronneberger2015u, olberg2019synthetic}.
    \item \textbf{Efficiency}: MMT can synthesize high-resolution images using a Transformer with tractable computation thanks to the proposed M-Swin Transformer blocks;
    \item \textbf{Interpretability}: The inherent attention maps inside MMT decoder indicate the amount of information coming from different input contrasts and regions for synthesizing every pixel in the output image.
\end{enumerate}}
\subsection{Multi-contrast Shifted Window-based Attention}
\label{sec:m-swin}
\begin{figure}[htbp]
  \centering
  \includegraphics[width=0.45\textwidth]{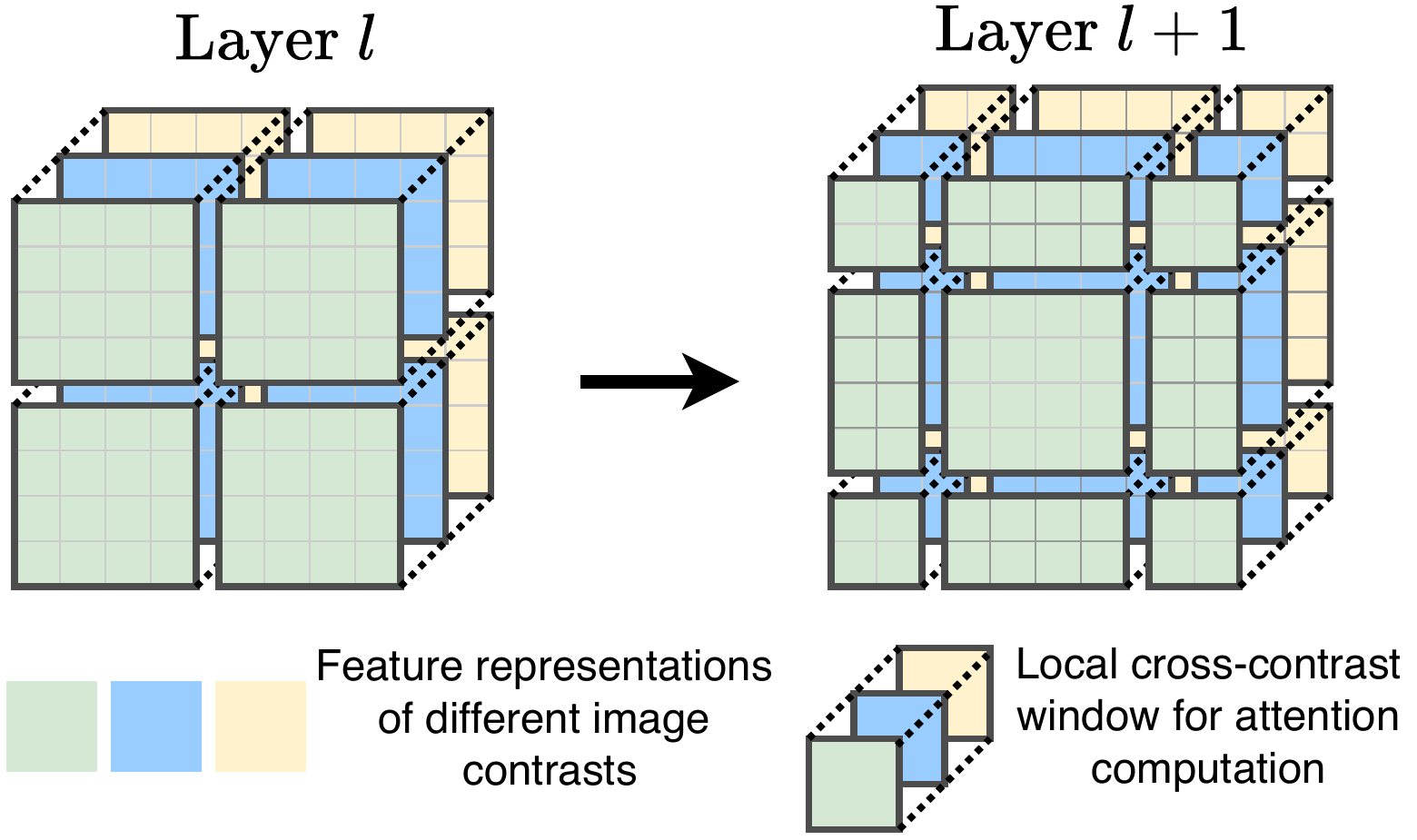}
  \caption{An illustration of the \textit{multi-contrast shifted window} approach (M-Swin) for attention computation. We compute attention within local cross-contrast windows to model inter- and intra-contrast dependencies. In layer $l$, a regular window partitioning is used, and in the next layer $l+1$, the window partitioning is shifted as in~\cite{swin}.} 
  \label{fig:mswin}
\end{figure}

\looseness -1 The core of MMT is the multi-contrast shifted window (M-Swin) based attention. \revise{M-Swin attention is a more general form of the window-based attention proposed in~\cite{swin}, which only processes single
modality data.} An illustration of M-Swin attention is shown in Fig.~\ref{fig:mswin}. We perform attention computation within local cross-contrast windows of size $W_h \times W_w$ to model both intra- and inter-contrast dependencies for accurate image synthesis. \revise{Specifically, within each cross-contrast window, every pixel attends to pixels of the same contrast to capture intra-contrast dependency, as well as pixels of other contrasts to capture inter-contrast dependency.} We use shifted window partitioning in successive blocks as in~\cite{swin} to enable connections between neighboring non-overlapping windows in the previous layer. Compared to global attention computation, this local window-based approach greatly reduces computational complexity for synthesizing high-resolution images as the complexity is quadratic with respect to the number of tokens~\cite{swin}. In addition, M-Swin attention can be computed regardless of the number of contrasts, which enables MMT to take any arbitrary subset of contrasts as input and generate the missing ones with only one model. \revise{The cross-contrast attention is also expected to improve the interpretability of the proposed model.}

We use MW-MHA and MSW-MHA to denote the M-Swin based multi-head attention (MHA) using regular and shifted window partitioning, respectively. 
\subsection{MMT Encoder}
\label{sec:encoder}
\begin{figure*}[htbp]
  \centering
  \includegraphics[width=1.0\textwidth]{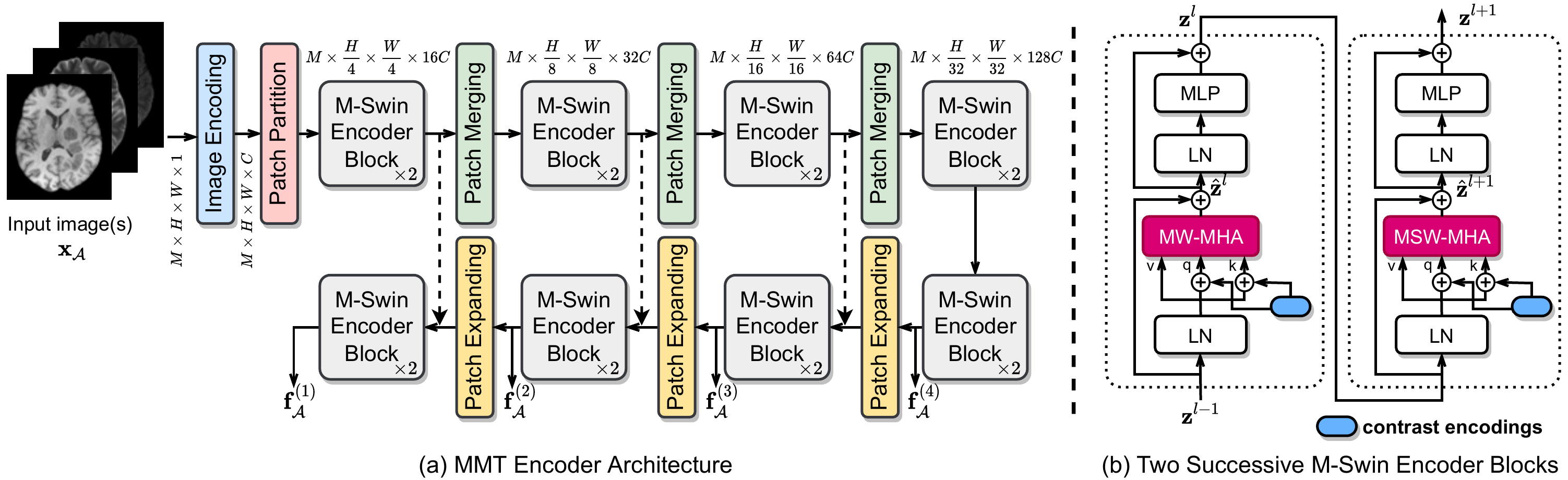}
  \caption{MMT Encoder architecture. MMT encoder performs joint encoding of multi-contrast input to capture inter- and intra-contrast dependencies and generates hierarchical representations of the input images at multiple scales. MW-MHA and MSW-MHA denote the M-Swin based multi-head attention using regular and shifted window partitioning, respectively.} 
  \label{fig:encoder}
\end{figure*}

\subsubsection{Overall architecture} The MMT encoder generates hierarchical representations of input images at multiple scales, which is shown to be beneficial in the medical imaging domain~\cite{ronneberger2015u, olberg2019synthetic}. It performs joint encoding of multi-contrast input to extract complementary information for accurate synthesis. The detailed architecture of the MMT encoder is shown in Fig.~\ref{fig:encoder} (a). The encoder is designed similarly to a U-Net~\cite{ronneberger2015u, cao2021swin} to generate multi-scale representations. The  $M$ input images of size $H\times W$ are first passed through separate image encoding blocks (described in Sec.~\ref{sec:cnn}) to project them to an arbitrary dimension $C$. The patch partition layer then splits each encoded image into non-overlapping patches and concatenates the features of each pixel. Each patch is considered as a ``token" for attention computation. We use a patch size of $4\times 4$, which results in $M\times \frac{H}{4}\times \frac{W}{4}$ patch tokens of feature dimension $4\times 4 \times C=16C$.

M-Swin encoder blocks are applied to the patch tokens to perform feature extraction. In the down-sampling path, two successive M-Swin encoder blocks are followed by a patch merging layer. The patch merging layer concatenates the features of each group of $2\times 2$ neighboring patches and applies a linear layer on the concatenated features, which results in $2\times$ reduction in spatial resolutions and $2\times$ increase in feature dimensions. Conversely, two successive M-Swin encoder blocks are followed by a patch-expanding layer in the up-sampling path. The patch-expanding layer first applies a linear layer to increase the feature dimensions by a factor of two. Then each patch token is split into $2\times 2$ neighboring tokens along the feature dimensions, which results in $2\times$ increase in spatial resolutions and $2\times$ reduction in feature dimensions. We concatenate the features from the down-sampling path with the up-sampled features produced by the patch-expanding layers to reduce the loss of spatial information, using a linear layer to retain the same feature dimension as the up-sampled features. 

The MMT encoder outputs the features $\mathbf{f}_\mathcal{A}^{(s)}=\{f_{a_i}^{(s)}\}_{i=1}^M$ at each stage $s$ in the up-sampling path $(s=1,2,3,4)$, which forms multi-scale representations of the input images.

\subsubsection{M-Swin encoder block} M-Swin encoder block is similar to a standard Transformer encoder block, except that the standard MHA is replaced by multi-contrast shifted window-based attention modules MW-MHA and MSW-MHA. Fig.~\ref{fig:encoder} (b) shows the detailed representation of the M-Swin encoder block setup. Consecutive M-Swin encoder blocks are formulated as:
\begin{equation}
\begin{aligned}
    &\hat{\textbf{z}}^l= \text{MW-MHA}(\text{LN}({\textbf{z}}^{l-1}))+{\textbf{z}}^{l-1},\\
    &{\textbf{z}}^l=\text{MLP}(\text{LN}(\hat{\textbf{z}}^l)) + \hat{\textbf{z}}^l,\\
    &\hat{\textbf{z}}^{l+1}= \text{MSW-MHA}(\text{LN}({\textbf{z}}^{l}))+{\textbf{z}}^{l},\\
    &{\textbf{z}}^{l+1}=\text{MLP}(\text{LN}(\hat{\textbf{z}}^{l+1})) + \hat{\textbf{z}}^{l+1},\\
\end{aligned}
\end{equation}
where LN is LayerNorm~\cite{ba2016layer}, MLP is a two-layer perceptron with GELU nonlinearity, and $\hat{\textbf{z}}^l$ and ${\textbf{z}}^l$ are the output features of the M(S)W-MHA module and the MLP module for block $l$ respectively. Since Transformers are permutation-invariant to the input sequence, we add \textit{contrast encodings} to inject contrast-specific information, which are learnable parameters for each contrast. Relative position bias is also added in attention computation as in~\cite{swin}.

\subsection{MMT Decoder}
\label{sec:decoder}
\begin{figure}[htbp]
  \centering
  \includegraphics[width=0.5\textwidth]{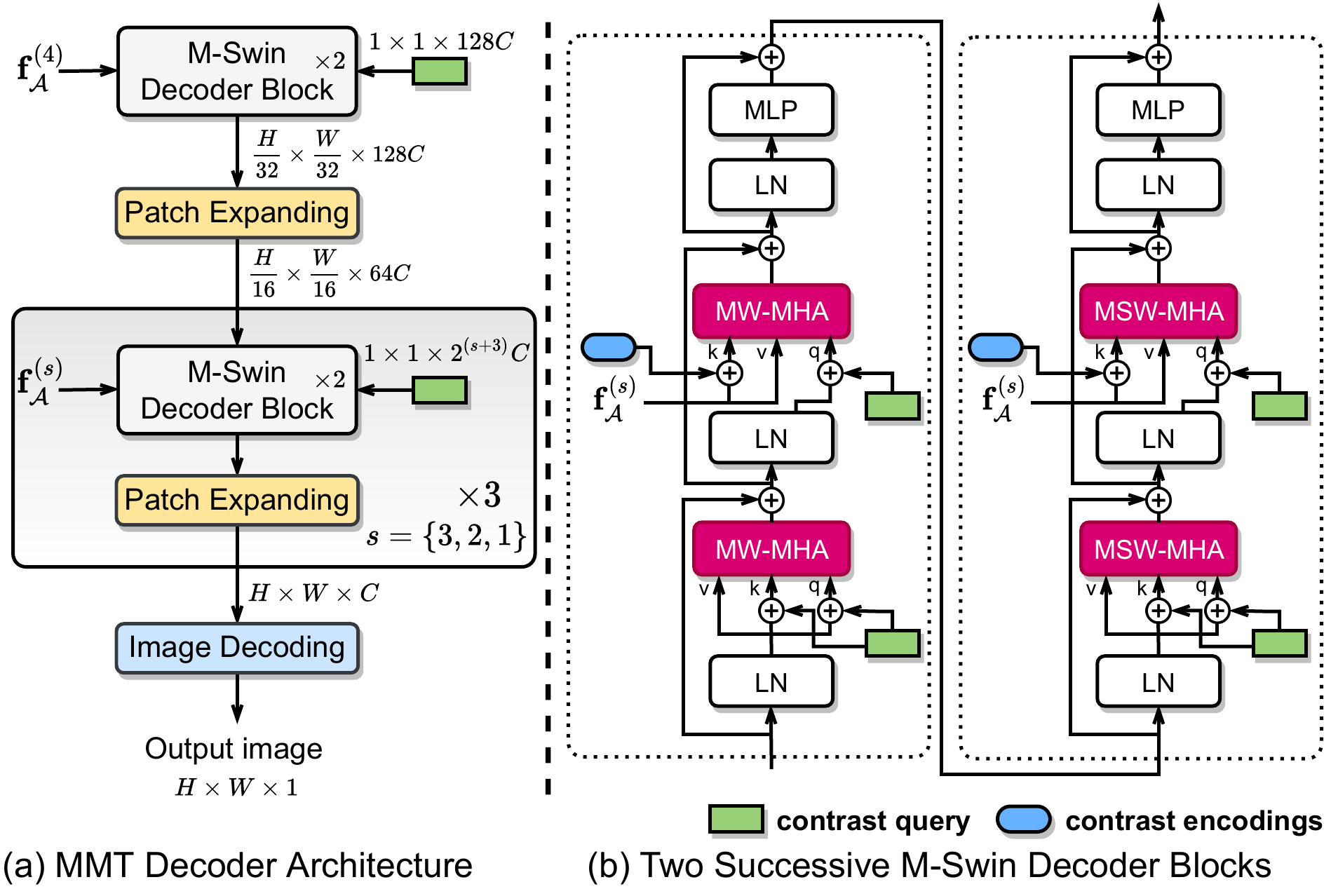}
  \caption{MMT Decoder architecture. The MMT decoder generates target output based on a \textit{contrast query}. The M-Swin decoder blocks progressively decode the encoder outputs at different scales and generate the desired output.} 
  \label{fig:decoder}
\end{figure}
\subsubsection{Overall Architecture} The MMT decoder works as a ``virtual scanner" that generates the target contrast based on the encoder outputs and the corresponding \textit{contrast query}. Contrast queries are learnable parameters that inform the decoder what contrast to synthesize and what information to decode from the encoder outputs. Fig.~\ref{fig:decoder} (a) shows the detailed representation of an MMT decoder. The M-Swin decoder progressively decodes the encoder outputs at different scales and generates the output image in a coarse-to-fine fashion, which allows it to consider both local and global image contexts for accurate image synthesis. Similar to the up-sampling path in the MMT encoder, two successive M-Swin decoder blocks are followed by a patch-expanding layer, which doubles the spatial resolutions and halves the feature dimensions. The last patch merging layer performs a $4\times$ up-sampling and restores the feature resolution to $H\times W$ by splitting each patch token into $4\times 4$ neighboring tokens along the feature dimensions, which reduces the feature dimension from $16C$ to $C$. Finally, an image decoding block (described in Sec.~\ref{sec:cnn}) is applied to produce the final image.

\subsubsection{M-Swin Decoder block}  
The detailed representation of the M-Swin decoder blocks is shown in Fig.~\ref{fig:decoder} (b). The M-Swin decoder block has a similar structure to the encoder block, except that an additional M(S)W-MHA layer decodes the outputs of the MMT encoder. The additional M(S)W-MHA layer takes the features of input contrasts as key $K$ and value $V$ and the feature of targeted contrast as query $Q$ in attention computation. It essentially compares the similarity between the input and target contrasts to compute the attention scores and then aggregates the features from input contrasts to produce the features of target contrasts using the attention scores as weights. Therefore, the attention scores in this layer provide a quantitative measurement of the amount of information flowing from different input contrasts and regions for synthesizing the output image, which makes MMT \textit{inherently interpretable}. We provide attention score analysis and visualization in Sec.~\ref{sec:att}.

\subsection{CNN Blocks}
\label{sec:cnn}
\begin{figure}[htbp]
  \centering
  \includegraphics[width=0.49\textwidth]{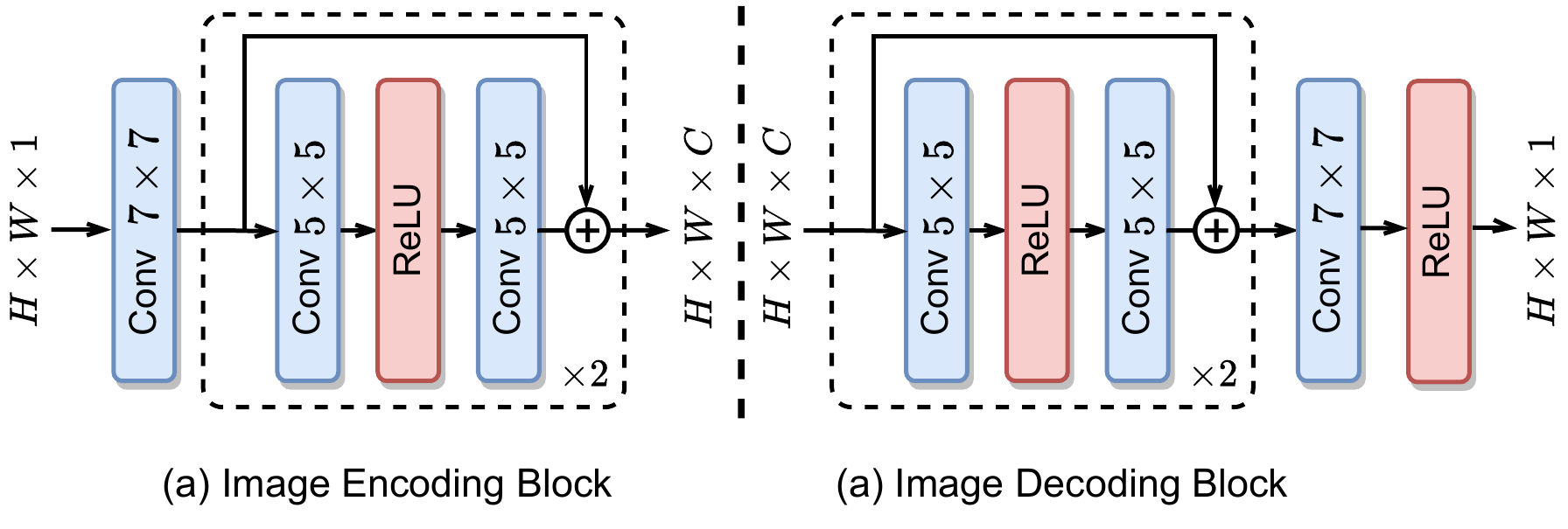}
  \caption{Architectures of the image encoding and decoding blocks. ``Conv $n \times n$" denotes a convolutional layer with kernel size $n\times n$ and ``ReLU" denotes a ReLU nonlinearity layer.} 
  \label{fig:cnn}
\end{figure}
CNNs are good at extracting image features. To combine the best of Transformer and CNNs, we use shallow CNN blocks for image encoding before feeding the images into Transformer blocks in the MMT encoder and final image decoding in the MMT decoder. The detailed architectures of image encoding and decoding blocks are shown in Fig.~\ref{fig:cnn}. \revise{We use separate encoding and decoding blocks for each contrast. We conduct an ablation study to validate the effectiveness of this design in Sec.~\ref{sec:ablation}.}

\subsection{Discriminator}
\label{sec:disc}
To further improve the perceptual quality of the synthetic images, we use CNN-based discriminators to adversarially train MMT. We employ the multi-scale discriminators proposed in~\cite{wang2018high} to guide MMT to produce realistic details and correct global structure. We train separate discriminators for each contrast in order to learn contrast-specific features. 

\subsection{Loss functions}
The loss function for training MMT includes three components: synthesis loss, reconstruction loss, and adversarial loss.

\subsubsection{Synthesis Loss} Synthesis loss measures the pixel-wise similarity between output images $\hat{\mathbf{x}}_{\mathcal{C}}=\{\hat{x}_{c_i}\}_{i=1}^{N}$ and the ground-truth images $\mathbf{x}_{\mathcal{C}}=\{{x}_{c_i}\}_{i=1}^{N}$, which is defined as:

\begin{equation}
    \mathcal{L}_s =  \frac{1}{N}\sum_{i=1}^{N}\mathop{\mathbb{E}}\left[{\| x_{c_i} - \hat{x}_{c_i}} \|_1\right].
    \label{eq:loss_s}
\end{equation}

Synthesis loss trains MMT to synthesize the missing contrasts when given the available contrasts accurately.

\subsubsection{Reconstruction Loss} MMT is expected to recover the input images $\mathbf{x}_{\mathcal{A}}=\{x_{a_i}\}_{i=1}^{M}$, when the decoder is queried with the contrast queries of input contrasts $\mathbf{q}_\mathcal{A}=\{q_{a_i}\}_{i=1}^{M}$. The reconstructed inputs $\hat{\mathbf{x}}_{\mathcal{A}}=\{\hat{x}_{a_i}\}_{i=1}^{M}$ are given by:
\begin{equation}
    \hat{x}_{a_i}=Dec(\mathbf{f}_\mathcal{A}; q_{a_i}),~a_i\in \mathcal{A}.    
\end{equation}
The reconstruction loss is defined as the $L_1$ distance between input images and reconstructed images:
\begin{equation}
    \mathcal{L}_r = \frac{1}{M} \sum_{i=1}^{M}\mathop{\mathbb{E}}\left[{\| x_{a_i} - \hat{x}_{a_i}} \|_1\right]. 
    \label{eq:loss_r}
\end{equation}
Reconstruction loss acts as a regularizer by forcing the model to reconstruct the inputs. It ensures that the feature representations generated by the MMT encoder preserve the information in the inputs.
\subsubsection{Adversarial Loss} Adversarial loss encourages MMT to generate realistic images to fool the discriminators. We use the LSGAN objective function~\cite{mao2017least}. The adversarial loss for training the discriminators is defined as:
\begin{equation}
\begin{split}
\mathcal{L}_{D} = \frac{1}{N} \sum_{i=1}^{N} \mathbb{E} \left[\left(D_{c_i}(\hat{x}_{c_i}) - Label_f\right)^2 \right. \\ + \left. \left(D_{c_i}({x}_{c_i}) - Label_r\right)^2\right].
\label{eq:adv_d}
\end{split}
\end{equation}

where $Label_f$ and $Label_r$ are the labels for fake and real images, respectively, and $D_{c_i}$ is the discriminator for contrast $c_i$. The adversarial loss for training MMT is defined as:
\begin{equation}
    \mathcal{L}_{adv} = \frac{1}{N} \sum_{i=1}^{N} \mathbb{E} \left[\left(D_{c_i}(\hat{x}_{c_i}) - Label_r\right)^2 \right]. 
    \label{eq:adv_g}
\end{equation}
We use label smoothing~\cite{NIPS2016_8a3363ab} to stabilize the training process where we sample the labels from uniform distributions instead of using fixed values of 0 or 1:
\begin{equation}
    Label_f \sim U(0, 0.1),~Label_r \sim U(0.9, 1). 
    \label{eq4}
\end{equation}
Adversarial learning between the discriminators and MMT network forces the distribution of the synthetic images to match that of real images for each contrast.
\subsubsection{Overall Loss} The overall loss of the generator $\mathcal{L}_G$ is defined as the weighted sum of the three loss components:
\begin{equation}
    \mathcal{L}_G = \lambda_r \mathcal{L}_r + \lambda_s \mathcal{L}_s + \lambda_{adv} \mathcal{L}_{adv},
    \label{eq9}
\end{equation}
where $\lambda_r$, $\lambda_s$ and $\lambda_{adv}$ are the weights for each loss term.
\section{Experiments and Results}
\subsection{Datasets}
\looseness -1 We evaluated the proposed methods on two multi-contrast brain MRI datasets: IXI~\cite{ixi} and BraTS 2021 (BraTS)~\cite{baid2021rsna}. The IXI dataset consists of 577 scans from normal, healthy subjects with three contrasts: T1, T2, and PD-weighted (PD). The images were neither skull-stripped nor pre-registered. We co-registered the T1 and PD images to T2 for each case using affine registration. We randomly selected 521, 28, and 28 cases for training, validation, and testing, respectively. We used the 90 middle axial slices and maintained the $256\times 256$ image size. The BraTS dataset consists of 1251 patient scans in patients with brain neoplasms with four contrasts: T1, post-contrast T1 weighted (T1Gd), T2-weighted (T2), and T2-FLAIR (FLAIR). All images were skull-stripped and co-registered to the same anatomical template. Two cases were excluded due to poor image quality. We randomly selected 1123, 63, and 63 cases for training, validation, and testing, respectively. We used the 80 middle axial slices and cropped the image size to $192\times160$ around the brain region. Mean normalization was performed for each image volume for both datasets.

\subsection{Evaluation Settings}

We evaluated the models using the peak signal-to-noise ratio (PSNR), structural similarity index (SSIM), and LPIPS~\cite{zhang2018unreasonable}, which captures the perceptual similarity between images. We compared MMT with two state-of-the-art CNN methods for missing data imputation: MILR~\cite{milr} and MMGAN~\cite{mmgan}. 
We considered two scenarios: 1) \textit{single missing contrast}, where only one contrast is missing, \ie, $N=1$; 2) \textit{random missing contrast}, where $N\in\{1, 2, \dots, K-1\}$ contrast(s) can be missing, which is the most general case. 

\subsection{Implementation Details}

\looseness -1 All experiments were performed on a server with 8 Nvidia Tesla V100 GPUs. We set the initial feature dimensions $C$ to 6. We trained the MMT models for 100 epochs using the AdamW optimizer~\cite{loshchilov2018decoupled}. We used an initial learning rate of $5\times 10^{-4}$ for the MMT generator and $1\times 10^{-4}$ for the discriminators with cosine learning rate schedulers. For batch size, we used 20 for the IXI dataset and 32 for the BraTS dataset. We set the window size $W_h \times W_w$ to $8 \times 8$ for the IXI dataset and $6 \times 5$ for the BraTS dataset. We set $\lambda_r = 5$, $\lambda_s = 20$ and $\lambda_{adv} = 0.1$ in Eq.~\ref{eq9}. \revise{The loss weights are chosen based on the qualitative and quantitative results on the validation sets.} For MILR, we implemented the algorithm as described in~\cite{milr}. For MMGAN, we used the source code provided by the authors and modified them accordingly for the single and random missing contrast scenarios. 
During training, we randomly sampled the set of available contrasts $\mathcal{A}$ at each iteration. 

\subsection{Handling variable inputs and outputs}
\looseness -1 \revise{
The proposed MMT network's forward call takes a variable length list as an argument that has all the available input contrasts. The method also accepts a list of indices present in the first argument and a list of indices the model must synthesize.
The encoder takes the available input images and the corresponding learned contrast encodings and generates the input contrasts' feature representations. Similarly, the decoder takes these encoder representations and learned contrast queries (for the missing contrasts) and synthesizes the images one at a time. At training time, the data loader generates a random combination of input/output contrasts for each batch so that the contrast encodings and the contrast queries are learned appropriately. The length of the input/output can be variable as the dot product operation in the attention computation can be done for any number of tokens, and the Multi-layer perceptron (MLP) used as part of the M-Swin blocks can also operate on a variable number of tokens.}

\begin{table}[t]
\centering
\caption{Summary of the quantitative metrics on different methods. $\uparrow$ / $\downarrow$ indicates that higher/lower values correspond to better image quality respectively. N is the number of missing contrasts. The best performance is in \textbf{bold} with $p < 0.005$}
\label{tab:main}
\setlength{\tabcolsep}{1.5pt}
\resizebox{\columnwidth}{!}{
\begin{tabular}{|c|c|c|l|c|c|c|}
\hline
Dataset & Model & N & Method & PSNR (dB) $\uparrow$ & SSIM $\uparrow$ & LPIPS $\downarrow$ \\ \hline
\multirow{9}{*}{IXI} & \multirow{3}{*}{Single} & \multirow{3}{*}{1} & MILR  & 34.69 $\pm$ 3.92 & 0.957 $\pm$ 0.031 & 0.125 $\pm$ 0.031 \\ \cline{4-7} 
 &  &  & MMGAN  & 35.79 $\pm$ 3.77 & 0.961 $\pm$ 0.027 & 0.111 $\pm$ 0.030 \\ \cline{4-7} 
 &  &  & \textbf{MMT} & \textbf{36.58 $\pm$ 3.69} & \textbf{0.963 $\pm$ 0.028} & \textbf{0.078 $\pm$ 0.026} \\ \cline{2-7} 
 & \multirow{6}{*}{Random} & \multirow{3}{*}{1} & MILR  & 34.89 $\pm$ 3.44 & 0.956 $\pm$ 0.032 & 0.127 $\pm$ 0.032 \\ \cline{4-7} 
 &  &  & MMGAN  & 35.45 $\pm$ 3.53 & 0.959 $\pm$ 0.028 & 0.107 $\pm$ 0.030 \\ \cline{4-7} 
 &  &  & \textbf{MMT} & \textbf{36.31 $\pm$ 3.64} & \textbf{0.961 $\pm$ 0.028} & \textbf{0.080 $\pm$ 0.025} \\ \cline{3-7} 
 &  & \multirow{3}{*}{2} & MILR  & 32.14 $\pm$ 4.01 & 0.928 $\pm$ 0.045 & 0.154 $\pm$ 0.047 \\ \cline{4-7} 
 &  &  & MMGAN  & 32.47 $\pm$ 4.15 & 0.932 $\pm$ 0.041 & 0.143 $\pm$ 0.049 \\ \cline{4-7} 
 &  &  & \textbf{MMT} & \textbf{33.35 $\pm$ 4.26} & \textbf{0.936 $\pm$ 0.041} & \textbf{0.108 $\pm$ 0.042} \\ \hline
\multirow{12}{*}{BraTS} & \multirow{3}{*}{Single} & \multirow{3}{*}{1} & MILR  & 27.30 $\pm$ 2.82 & 0.927 $\pm$ 0.033 & 0.121 $\pm$ 0.047 \\ \cline{4-7} 
 &  &  & MMGAN  & 27.15 $\pm$ 2.52 & 0.924 $\pm$ 0.032 & 0.130 $\pm$ 0.044 \\ \cline{4-7} 
 &  &  & \textbf{MMT} & \textbf{27.74 $\pm$ 2.83} & \textbf{0.931 $\pm$ 0.031} & \textbf{0.100 $\pm$ 0.041} \\ \cline{2-7} 
 & \multirow{9}{*}{Random} & \multirow{3}{*}{1} & MILR  & 27.20 $\pm$ 2.77 & 0.926 $\pm$ 0.033 & 0.123 $\pm$ 0.047 \\ \cline{4-7} 
 &  &  & MMGAN  & 27.00 $\pm$ 2.50 & 0.922 $\pm$ 0.034 & 0.131 $\pm$ 0.045 \\ \cline{4-7} 
 &  &  & \textbf{MMT} & \textbf{27.87 $\pm$ 2.77} & \textbf{0.932 $\pm$ 0.031} & \textbf{0.104 $\pm$ 0.044} \\ \cline{3-7} 
 &  & \multirow{3}{*}{2} & MILR  & 26.45 $\pm$ 2.76 & 0.917 $\pm$ 0.037 & 0.131 $\pm$ 0.049 \\ \cline{4-7} 
 &  &  & MMGAN  & 26.21 $\pm$ 2.56 & 0.913 $\pm$ 0.037 & 0.141 $\pm$ 0.046 \\ \cline{4-7} 
 &  &  & \textbf{MMT} & \textbf{27.02 $\pm$ 2.69} & \textbf{0.922 $\pm$ 0.034} & \textbf{0.116 $\pm$ 0.047} \\ \cline{3-7} 
 &  & \multirow{3}{*}{3} & MILR  & 25.24 $\pm$ 2.72 & 0.898 $\pm$ 0.042 & 0.146 $\pm$ 0.052 \\ \cline{4-7} 
 &  &  & MMGAN  & 25.05 $\pm$ 2.58 & 0.894 $\pm$ 0.041 & 0.158 $\pm$ 0.049 \\ \cline{4-7} 
 &  &  & \textbf{MMT} & \textbf{25.63 $\pm$ 2.59} & \textbf{0.902 $\pm$ 0.040} & \textbf{0.136 $\pm$ 0.050} \\ \hline
\end{tabular}
}
\centering
\end{table}
\begin{table}[h]
\centering
\caption{\revise{Quantitative performance of the proposed MMT random model along with the other comparative methods on the IXI dataset for all input combinations.}}
\label{tab:random_ixi}
\setlength{\tabcolsep}{3pt}
{
\begin{subtable}[h]{0.45\textwidth}
\caption{\revise{MILR}}
\begin{tabular}{|ccc|ccc|}
\hline
\multicolumn{3}{|c|}{\textbf{Input}} & \multicolumn{3}{c|}{\textbf{Metrics {[}PSNR (dB) $\uparrow$ / SSIM $\uparrow$ / LPIPS $\downarrow${]}}} \\ \hline
\multicolumn{1}{|c|}{\textbf{T1}} & \multicolumn{1}{c|}{\textbf{T2}} & \textbf{PD} & \multicolumn{1}{c|}{\textbf{T1}} & \multicolumn{1}{c|}{\textbf{T2}} & \textbf{PD} \\ \hline
\multicolumn{1}{|c|}{\ding{55}} & \multicolumn{1}{c|}{\ding{55}} & \ding{51} & \multicolumn{1}{c|}{32.96/0.917/0.140} & \multicolumn{1}{c|}{31.65/0.968/0.131} & - \\ \hline
\multicolumn{1}{|c|}{\ding{55}} & \multicolumn{1}{c|}{\ding{51}} & \ding{55} & \multicolumn{1}{c|}{32.31/0.913/0.145} & \multicolumn{1}{c|}{-} & 37.60/0.970/0.120 \\ \hline
\multicolumn{1}{|c|}{\ding{51}} & \multicolumn{1}{c|}{\ding{55}} & \ding{55} & \multicolumn{1}{c|}{-} & \multicolumn{1}{c|}{25.93/0.897/0.201} & 32.39/0.906/0.188 \\ \hline
\multicolumn{1}{|c|}{\ding{55}} & \multicolumn{1}{c|}{\ding{51}} & \ding{51} & \multicolumn{1}{c|}{33.19/0.925/0.135} & \multicolumn{1}{c|}{-} & - \\ \hline
\multicolumn{1}{|c|}{\ding{51}} & \multicolumn{1}{c|}{\ding{55}} & \ding{51} & \multicolumn{1}{c|}{-} & \multicolumn{1}{c|}{32.22/0.972/0.129} & - \\ \hline
\multicolumn{1}{|c|}{\ding{51}} & \multicolumn{1}{c|}{\ding{51}} & \ding{55} & \multicolumn{1}{c|}{-} & \multicolumn{1}{c|}{-} & 38.97/0.974/0.116 \\ \hline
\end{tabular}
\end{subtable}
}
{
\begin{subtable}[h]{0.45\textwidth}
\caption{\revise{MMGAN}}
\begin{tabular}{|ccc|ccc|}
\hline
\multicolumn{3}{|c|}{\textbf{Input}} & \multicolumn{3}{c|}{\textbf{Metrics {[}PSNR (dB) $\uparrow$ / SSIM $\uparrow$ / LPIPS $\downarrow${]}}} \\ \hline
\multicolumn{1}{|c|}{\textbf{T1}} & \multicolumn{1}{c|}{\textbf{T2}} & \textbf{PD} & \multicolumn{1}{c|}{\textbf{T1}} & \multicolumn{1}{c|}{\textbf{T2}} & \textbf{PD} \\ \hline
\multicolumn{1}{|c|}{\ding{55}} & \multicolumn{1}{c|}{\ding{55}} & \ding{51} & \multicolumn{1}{c|}{33.77/0.925/0.135} & \multicolumn{1}{c|}{31.82/0.968/0.110} & - \\ \hline
\multicolumn{1}{|c|}{\ding{55}} & \multicolumn{1}{c|}{\ding{51}} & \ding{55} & \multicolumn{1}{c|}{33.34/0.923/0.135} & \multicolumn{1}{c|}{-} & 38.12/0.970/0.100 \\ \hline
\multicolumn{1}{|c|}{\ding{51}} & \multicolumn{1}{c|}{\ding{55}} & \ding{55} & \multicolumn{1}{c|}{-} & \multicolumn{1}{c|}{25.72/0.900/0.197} & 32.08/0.907/0.185 \\ \hline
\multicolumn{1}{|c|}{\ding{55}} & \multicolumn{1}{c|}{\ding{51}} & \ding{51} & \multicolumn{1}{c|}{34.36/0.934/0.124} & \multicolumn{1}{c|}{-} & - \\ \hline
\multicolumn{1}{|c|}{\ding{51}} & \multicolumn{1}{c|}{\ding{55}} & \ding{51} & \multicolumn{1}{c|}{-} & \multicolumn{1}{c|}{32.75/0.973/0.111} & - \\ \hline
\multicolumn{1}{|c|}{\ding{51}} & \multicolumn{1}{c|}{\ding{51}} & \ding{55} & \multicolumn{1}{c|}{-} & \multicolumn{1}{c|}{-} & 39.76/0.975/0.094 \\ \hline
\end{tabular}
\end{subtable}
}
{
\begin{subtable}[h]{0.45\textwidth}
\caption{\textbf{MMT}}
\begin{tabular}{|ccc|ccc|}
\hline
\multicolumn{3}{|c|}{\textbf{Input}} & \multicolumn{3}{c|}{\textbf{Metrics {[}PSNR (dB) $\uparrow$ / SSIM $\uparrow$ / LPIPS $\downarrow${]}}} \\ \hline
\multicolumn{1}{|c|}{\textbf{T1}} & \multicolumn{1}{c|}{\textbf{T2}} & \textbf{PD} & \multicolumn{1}{c|}{\textbf{T1}} & \multicolumn{1}{c|}{\textbf{T2}} & \textbf{PD} \\ \hline
\multicolumn{1}{|c|}{\ding{55}} & \multicolumn{1}{c|}{\ding{55}} & \ding{51} & \multicolumn{1}{c|}{34.02/0.928/0.104} & \multicolumn{1}{c|}{33.01/0.971/0.078} & - \\ \hline
\multicolumn{1}{|c|}{\ding{55}} & \multicolumn{1}{c|}{\ding{51}} & \ding{55} & \multicolumn{1}{c|}{33.71/0.926/0.104} & \multicolumn{1}{c|}{-} & 39.50/0.974/0.073 \\ \hline
\multicolumn{1}{|c|}{\ding{51}} & \multicolumn{1}{c|}{\ding{55}} & \ding{55} & \multicolumn{1}{c|}{-} & \multicolumn{1}{c|}{26.70/0.906/0.149} & 33.22/0.913/0.143 \\ \hline
\multicolumn{1}{|c|}{\ding{55}} & \multicolumn{1}{c|}{\ding{51}} & \ding{51} & \multicolumn{1}{c|}{34.52/0.935/0.096} & \multicolumn{1}{c|}{-} & - \\ \hline
\multicolumn{1}{|c|}{\ding{51}} & \multicolumn{1}{c|}{\ding{55}} & \ding{51} & \multicolumn{1}{c|}{-} & \multicolumn{1}{c|}{34.16/0.975/0.074} & - \\ \hline
\multicolumn{1}{|c|}{\ding{51}} & \multicolumn{1}{c|}{\ding{51}} & \ding{55} & \multicolumn{1}{c|}{-} & \multicolumn{1}{c|}{-} & 40.69/0.977/0.069 \\ \hline
\end{tabular}
\end{subtable}
}
\end{table}

\subsection{Evaluation Results}

For each method and each dataset, we trained two models for the single and random missing contrast scenario, respectively, and we use \textit{``single"} and \textit{``random"} to refer to the models trained in the two scenarios.
Table~\ref{tab:main} summarizes the quantitative results of different methods on the test sets. We report the average performance across all output contrasts and all input-output combinations when $N$ contrasts are missing. Overall, the performance of all methods decreases as $N$ increases because less information is available in the inputs. The performance of random models is similar to single models when $N=1$. MMT performs significantly better than MILR~\cite{milr} and MMGAN~\cite{mmgan} in all metrics and all scenarios on both IXI and BraTS datasets, based on Wilcoxon signed-rank test with $p < 0.005$. The LPIPS scores of MMT are much lower than MILR and MMGAN, which indicates that the outputs of MMT have improved perceptual quality.

\begin{table}[htbp]
\centering
\caption{\revise{Quantitative performance of the proposed MMT random model along with the other comparative methods on the BraTS dataset for all input combinations.}}
\label{tab:random_brats}
\setlength{\tabcolsep}{1.1pt}
\resizebox{\columnwidth}{!}{
\begin{subtable}[h]{0.6\textwidth}
\caption{\revise{MILR}}
\begin{tabular}{|cccc|cccc|}
\hline
\multicolumn{4}{|c|}{\textbf{Input}} & \multicolumn{4}{c|}{\textbf{Metrics {[}PSNR (dB) $\uparrow$ / SSIM $\uparrow$ / LPIPS $\downarrow${]}}} \\ \hline
\multicolumn{1}{|c|}{\textbf{T1}} & \multicolumn{1}{c|}{\textbf{T1Gd}} & \multicolumn{1}{c|}{\textbf{T2}} & \textbf{FLAIR} & \multicolumn{1}{c|}{\textbf{T1}} & \multicolumn{1}{c|}{\textbf{T1Gd}} & \multicolumn{1}{c|}{\textbf{T2}} & \textbf{FLAIR} \\ \hline
\multicolumn{1}{|c|}{\ding{55}} & \multicolumn{1}{c|}{\ding{55}} & \multicolumn{1}{c|}{\ding{55}} & \ding{51} & \multicolumn{1}{c|}{23.85/0.894/0.146} & \multicolumn{1}{c|}{26.51/0.899/0.176} & \multicolumn{1}{c|}{24.45/0.900/0.125} & - \\ \hline
\multicolumn{1}{|c|}{\ding{55}} & \multicolumn{1}{c|}{\ding{55}} & \multicolumn{1}{c|}{\ding{51}} & \ding{55} & \multicolumn{1}{c|}{24.09/0.910/0.133} & \multicolumn{1}{c|}{27.09/0.912/0.156} & \multicolumn{1}{c|}{-} & 24.31/0.885/0.152 \\ \hline
\multicolumn{1}{|c|}{\ding{55}} & \multicolumn{1}{c|}{\ding{51}} & \multicolumn{1}{c|}{\ding{55}} & \ding{55} & \multicolumn{1}{c|}{27.15/0.926/0.128} & \multicolumn{1}{c|}{-} & \multicolumn{1}{c|}{24.55/0.899/0.137} & 23.63/0.859/0.173 \\ \hline
\multicolumn{1}{|c|}{\ding{51}} & \multicolumn{1}{c|}{\ding{55}} & \multicolumn{1}{c|}{\ding{55}} & \ding{55} & \multicolumn{1}{c|}{-} & \multicolumn{1}{c|}{28.57/0.926/0.146} & \multicolumn{1}{c|}{24.91/0.910/0.120} & 23.80/0.867/0.164 \\ \hline
\multicolumn{1}{|c|}{\ding{55}} & \multicolumn{1}{c|}{\ding{55}} & \multicolumn{1}{c|}{\ding{51}} & \ding{51} & \multicolumn{1}{c|}{24.86/0.919/0.128} & \multicolumn{1}{c|}{27.57/0.918/0.151} & \multicolumn{1}{c|}{-} & - \\ \hline
\multicolumn{1}{|c|}{\ding{55}} & \multicolumn{1}{c|}{\ding{51}} & \multicolumn{1}{c|}{\ding{55}} & \ding{51} & \multicolumn{1}{c|}{27.51/0.934/0.121} & \multicolumn{1}{c|}{-} & \multicolumn{1}{c|}{25.84/0.924/0.109} & - \\ \hline
\multicolumn{1}{|c|}{\ding{55}} & \multicolumn{1}{c|}{\ding{51}} & \multicolumn{1}{c|}{\ding{51}} & \ding{55} & \multicolumn{1}{c|}{27.54/0.936/0.118} & \multicolumn{1}{c|}{-} & \multicolumn{1}{c|}{-} & 25.14/0.895/0.146 \\ \hline
\multicolumn{1}{|c|}{\ding{51}} & \multicolumn{1}{c|}{\ding{55}} & \multicolumn{1}{c|}{\ding{55}} & \ding{51} & \multicolumn{1}{c|}{-} & \multicolumn{1}{c|}{28.90/0.929/0.142} & \multicolumn{1}{c|}{25.96/0.928/0.103} & - \\ \hline
\multicolumn{1}{|c|}{\ding{51}} & \multicolumn{1}{c|}{\ding{55}} & \multicolumn{1}{c|}{\ding{51}} & \ding{55} & \multicolumn{1}{c|}{-} & \multicolumn{1}{c|}{29.25/0.934/0.136} & \multicolumn{1}{c|}{-} & 25.32/0.898/0.144 \\ \hline
\multicolumn{1}{|c|}{\ding{51}} & \multicolumn{1}{c|}{\ding{51}} & \multicolumn{1}{c|}{\ding{55}} & \ding{55} & \multicolumn{1}{c|}{-} & \multicolumn{1}{c|}{-} & \multicolumn{1}{c|}{25.39/0.918/0.114} & 24.22/0.875/0.161 \\ \hline
\multicolumn{1}{|c|}{\ding{55}} & \multicolumn{1}{c|}{\ding{51}} & \multicolumn{1}{c|}{\ding{51}} & \ding{51} & \multicolumn{1}{c|}{27.67/0.939/0.116} & \multicolumn{1}{c|}{-} & \multicolumn{1}{c|}{-} & - \\ \hline
\multicolumn{1}{|c|}{\ding{51}} & \multicolumn{1}{c|}{\ding{55}} & \multicolumn{1}{c|}{\ding{51}} & \ding{51} & \multicolumn{1}{c|}{-} & \multicolumn{1}{c|}{29.52/0.936/0.133} & \multicolumn{1}{c|}{-} & - \\ \hline
\multicolumn{1}{|c|}{\ding{51}} & \multicolumn{1}{c|}{\ding{51}} & \multicolumn{1}{c|}{\ding{55}} & \ding{51} & \multicolumn{1}{c|}{-} & \multicolumn{1}{c|}{-} & \multicolumn{1}{c|}{26.30/0.933/0.099} & - \\ \hline
\multicolumn{1}{|c|}{\ding{51}} & \multicolumn{1}{c|}{\ding{51}} & \multicolumn{1}{c|}{\ding{51}} & \ding{55} & \multicolumn{1}{c|}{-} & \multicolumn{1}{c|}{-} & \multicolumn{1}{c|}{-} & 25.53/0.901/0.142 \\ \hline
\end{tabular}
\end{subtable}
}

\resizebox{\columnwidth}{!}{
\begin{subtable}[h]{0.6\textwidth}
\caption{\revise{MMGAN}}
\begin{tabular}{|cccc|cccc|}
\hline
\multicolumn{4}{|c|}{\textbf{Input}} & \multicolumn{4}{c|}{\textbf{Metrics {[}PSNR (dB) $\uparrow$ / SSIM $\uparrow$ / LPIPS $\downarrow${]}}} \\ \hline
\multicolumn{1}{|c|}{\textbf{T1}} & \multicolumn{1}{c|}{\textbf{T1Gd}} & \multicolumn{1}{c|}{\textbf{T2}} & \textbf{FLAIR} & \multicolumn{1}{c|}{\textbf{T1}} & \multicolumn{1}{c|}{\textbf{T1Gd}} & \multicolumn{1}{c|}{\textbf{T2}} & \textbf{FLAIR} \\ \hline
\multicolumn{1}{|c|}{\ding{55}} & \multicolumn{1}{c|}{\ding{55}} & \multicolumn{1}{c|}{\ding{55}} & \ding{51} & \multicolumn{1}{c|}{23.72/0.885/0.162} & \multicolumn{1}{c|}{26.30/0.895/0.190} & \multicolumn{1}{c|}{24.31/0.894/0.140} & - \\ \hline
\multicolumn{1}{|c|}{\ding{55}} & \multicolumn{1}{c|}{\ding{55}} & \multicolumn{1}{c|}{\ding{51}} & \ding{55} & \multicolumn{1}{c|}{23.94/0.906/0.149} & \multicolumn{1}{c|}{26.69/0.908/0.167} & \multicolumn{1}{c|}{-} & 24.44/0.879/0.158 \\ \hline
\multicolumn{1}{|c|}{\ding{55}} & \multicolumn{1}{c|}{\ding{51}} & \multicolumn{1}{c|}{\ding{55}} & \ding{55} & \multicolumn{1}{c|}{26.12/0.915/0.145} & \multicolumn{1}{c|}{-} & \multicolumn{1}{c|}{24.49/0.897/0.146} & 23.63/0.857/0.183 \\ \hline
\multicolumn{1}{|c|}{\ding{51}} & \multicolumn{1}{c|}{\ding{55}} & \multicolumn{1}{c|}{\ding{55}} & \ding{55} & \multicolumn{1}{c|}{-} & \multicolumn{1}{c|}{28.02/0.920/0.156} & \multicolumn{1}{c|}{25.00/0.910/0.131} & 24.02/0.868/0.171 \\ \hline
\multicolumn{1}{|c|}{\ding{55}} & \multicolumn{1}{c|}{\ding{55}} & \multicolumn{1}{c|}{\ding{51}} & \ding{51} & \multicolumn{1}{c|}{24.63/0.915/0.140} & \multicolumn{1}{c|}{27.20/0.914/0.162} & \multicolumn{1}{c|}{-} & - \\ \hline
\multicolumn{1}{|c|}{\ding{55}} & \multicolumn{1}{c|}{\ding{51}} & \multicolumn{1}{c|}{\ding{55}} & \ding{51} & \multicolumn{1}{c|}{26.79/0.925/0.132} & \multicolumn{1}{c|}{-} & \multicolumn{1}{c|}{25.81/0.921/0.121} & - \\ \hline
\multicolumn{1}{|c|}{\ding{55}} & \multicolumn{1}{c|}{\ding{51}} & \multicolumn{1}{c|}{\ding{51}} & \ding{55} & \multicolumn{1}{c|}{26.73/0.930/0.131} & \multicolumn{1}{c|}{-} & \multicolumn{1}{c|}{-} & 25.20/0.890/0.153 \\ \hline
\multicolumn{1}{|c|}{\ding{51}} & \multicolumn{1}{c|}{\ding{55}} & \multicolumn{1}{c|}{\ding{55}} & \ding{51} & \multicolumn{1}{c|}{-} & \multicolumn{1}{c|}{28.33/0.922/0.153} & \multicolumn{1}{c|}{25.97/0.926/0.115} & - \\ \hline
\multicolumn{1}{|c|}{\ding{51}} & \multicolumn{1}{c|}{\ding{55}} & \multicolumn{1}{c|}{\ding{51}} & \ding{55} & \multicolumn{1}{c|}{-} & \multicolumn{1}{c|}{28.54/0.927/0.146} & \multicolumn{1}{c|}{-} & 25.50/0.895/0.149 \\ \hline
\multicolumn{1}{|c|}{\ding{51}} & \multicolumn{1}{c|}{\ding{51}} & \multicolumn{1}{c|}{\ding{55}} & \ding{55} & \multicolumn{1}{c|}{-} & \multicolumn{1}{c|}{-} & \multicolumn{1}{c|}{25.44/0.917/0.124} & 24.45/0.875/0.166 \\ \hline
\multicolumn{1}{|c|}{\ding{55}} & \multicolumn{1}{c|}{\ding{51}} & \multicolumn{1}{c|}{\ding{51}} & \ding{51} & \multicolumn{1}{c|}{27.16/0.935/0.127} & \multicolumn{1}{c|}{-} & \multicolumn{1}{c|}{-} & - \\ \hline
\multicolumn{1}{|c|}{\ding{51}} & \multicolumn{1}{c|}{\ding{55}} & \multicolumn{1}{c|}{\ding{51}} & \ding{51} & \multicolumn{1}{c|}{-} & \multicolumn{1}{c|}{28.97/0.930/0.144} & \multicolumn{1}{c|}{-} & - \\ \hline
\multicolumn{1}{|c|}{\ding{51}} & \multicolumn{1}{c|}{\ding{51}} & \multicolumn{1}{c|}{\ding{55}} & \ding{51} & \multicolumn{1}{c|}{-} & \multicolumn{1}{c|}{-} & \multicolumn{1}{c|}{26.40/0.931/0.110} & - \\ \hline
\multicolumn{1}{|c|}{\ding{51}} & \multicolumn{1}{c|}{\ding{51}} & \multicolumn{1}{c|}{\ding{51}} & \ding{55} & \multicolumn{1}{c|}{-} & \multicolumn{1}{c|}{-} & \multicolumn{1}{c|}{-} & 25.79/0.899/0.145 \\ \hline
\end{tabular}
\end{subtable}
}

\resizebox{\columnwidth}{!}{
\begin{subtable}[h]{0.6\textwidth}
\caption{\textbf{MMT}}
\begin{tabular}{|cccc|cccc|}
\hline
\multicolumn{4}{|c|}{\textbf{Input}} & \multicolumn{4}{c|}{\textbf{Metrics {[}PSNR (dB) $\uparrow$ / SSIM $\uparrow$ / LPIPS $\downarrow${]}}} \\ \hline
\multicolumn{1}{|c|}{\textbf{T1}} & \multicolumn{1}{c|}{\textbf{T1Gd}} & \multicolumn{1}{c|}{\textbf{T2}} & \textbf{FLAIR} & \multicolumn{1}{c|}{\textbf{T1}} & \multicolumn{1}{c|}{\textbf{T1Gd}} & \multicolumn{1}{c|}{\textbf{T2}} & \textbf{FLAIR} \\ \hline
\multicolumn{1}{|c|}{\ding{55}} & \multicolumn{1}{c|}{\ding{55}} & \multicolumn{1}{c|}{\ding{55}} & \ding{51} & \multicolumn{1}{c|}{24.09/0.889/0.128} & \multicolumn{1}{c|}{26.86/0.902/0.172} & \multicolumn{1}{c|}{24.84/0.902/0.124} & - \\ \hline
\multicolumn{1}{|c|}{\ding{55}} & \multicolumn{1}{c|}{\ding{55}} & \multicolumn{1}{c|}{\ding{51}} & \ding{55} & \multicolumn{1}{c|}{24.38/0.909/0.114} & \multicolumn{1}{c|}{27.18/0.913/0.155} & \multicolumn{1}{c|}{-} & 25.00/0.891/0.141 \\ \hline
\multicolumn{1}{|c|}{\ding{55}} & \multicolumn{1}{c|}{\ding{51}} & \multicolumn{1}{c|}{\ding{55}} & \ding{55} & \multicolumn{1}{c|}{27.05/0.924/0.098} & \multicolumn{1}{c|}{-} & \multicolumn{1}{c|}{25.25/0.906/0.130} & 24.37/0.871/0.163 \\ \hline
\multicolumn{1}{|c|}{\ding{51}} & \multicolumn{1}{c|}{\ding{55}} & \multicolumn{1}{c|}{\ding{55}} & \ding{55} & \multicolumn{1}{c|}{-} & \multicolumn{1}{c|}{28.70/0.927/0.141} & \multicolumn{1}{c|}{25.38/0.916/0.117} & 24.56/0.879/0.155 \\ \hline
\multicolumn{1}{|c|}{\ding{55}} & \multicolumn{1}{c|}{\ding{55}} & \multicolumn{1}{c|}{\ding{51}} & \ding{51} & \multicolumn{1}{c|}{25.43/0.922/0.101} & \multicolumn{1}{c|}{27.93/0.922/0.147} & \multicolumn{1}{c|}{-} & - \\ \hline
\multicolumn{1}{|c|}{\ding{55}} & \multicolumn{1}{c|}{\ding{51}} & \multicolumn{1}{c|}{\ding{55}} & \ding{51} & \multicolumn{1}{c|}{27.58/0.932/0.088} & \multicolumn{1}{c|}{-} & \multicolumn{1}{c|}{26.60/0.929/0.097} & - \\ \hline
\multicolumn{1}{|c|}{\ding{55}} & \multicolumn{1}{c|}{\ding{51}} & \multicolumn{1}{c|}{\ding{51}} & \ding{55} & \multicolumn{1}{c|}{27.78/0.938/0.084} & \multicolumn{1}{c|}{-} & \multicolumn{1}{c|}{-} & 26.04/0.905/0.132 \\ \hline
\multicolumn{1}{|c|}{\ding{51}} & \multicolumn{1}{c|}{\ding{55}} & \multicolumn{1}{c|}{\ding{55}} & \ding{51} & \multicolumn{1}{c|}{-} & \multicolumn{1}{c|}{29.27/0.932/0.134} & \multicolumn{1}{c|}{26.74/0.934/0.092} & - \\ \hline
\multicolumn{1}{|c|}{\ding{51}} & \multicolumn{1}{c|}{\ding{55}} & \multicolumn{1}{c|}{\ding{51}} & \ding{55} & \multicolumn{1}{c|}{-} & \multicolumn{1}{c|}{29.55/0.936/0.128} & \multicolumn{1}{c|}{-} & 26.20/0.909/0.130 \\ \hline
\multicolumn{1}{|c|}{\ding{51}} & \multicolumn{1}{c|}{\ding{51}} & \multicolumn{1}{c|}{\ding{55}} & \ding{55} & \multicolumn{1}{c|}{-} & \multicolumn{1}{c|}{-} & \multicolumn{1}{c|}{26.13/0.926/0.108} & 25.09/0.888/0.151 \\ \hline
\multicolumn{1}{|c|}{\ding{55}} & \multicolumn{1}{c|}{\ding{51}} & \multicolumn{1}{c|}{\ding{51}} & \ding{51} & \multicolumn{1}{c|}{28.06/0.941/0.081} & \multicolumn{1}{c|}{-} & \multicolumn{1}{c|}{-} & - \\ \hline
\multicolumn{1}{|c|}{\ding{51}} & \multicolumn{1}{c|}{\ding{55}} & \multicolumn{1}{c|}{\ding{51}} & \ding{51} & \multicolumn{1}{c|}{-} & \multicolumn{1}{c|}{29.74/0.939/0.120} & \multicolumn{1}{c|}{-} & - \\ \hline
\multicolumn{1}{|c|}{\ding{51}} & \multicolumn{1}{c|}{\ding{51}} & \multicolumn{1}{c|}{\ding{55}} & \ding{51} & \multicolumn{1}{c|}{-} & \multicolumn{1}{c|}{-} & \multicolumn{1}{c|}{27.13/0.939/0.086} & - \\ \hline
\multicolumn{1}{|c|}{\ding{51}} & \multicolumn{1}{c|}{\ding{51}} & \multicolumn{1}{c|}{\ding{51}} & \ding{55} & \multicolumn{1}{c|}{-} & \multicolumn{1}{c|}{-} & \multicolumn{1}{c|}{-} & 26.30/0.911/0.125 \\ \hline
\end{tabular}
\end{subtable}
}
\end{table}
\begin{figure}[h]
  \centering
  \includegraphics[width=1.0\linewidth]{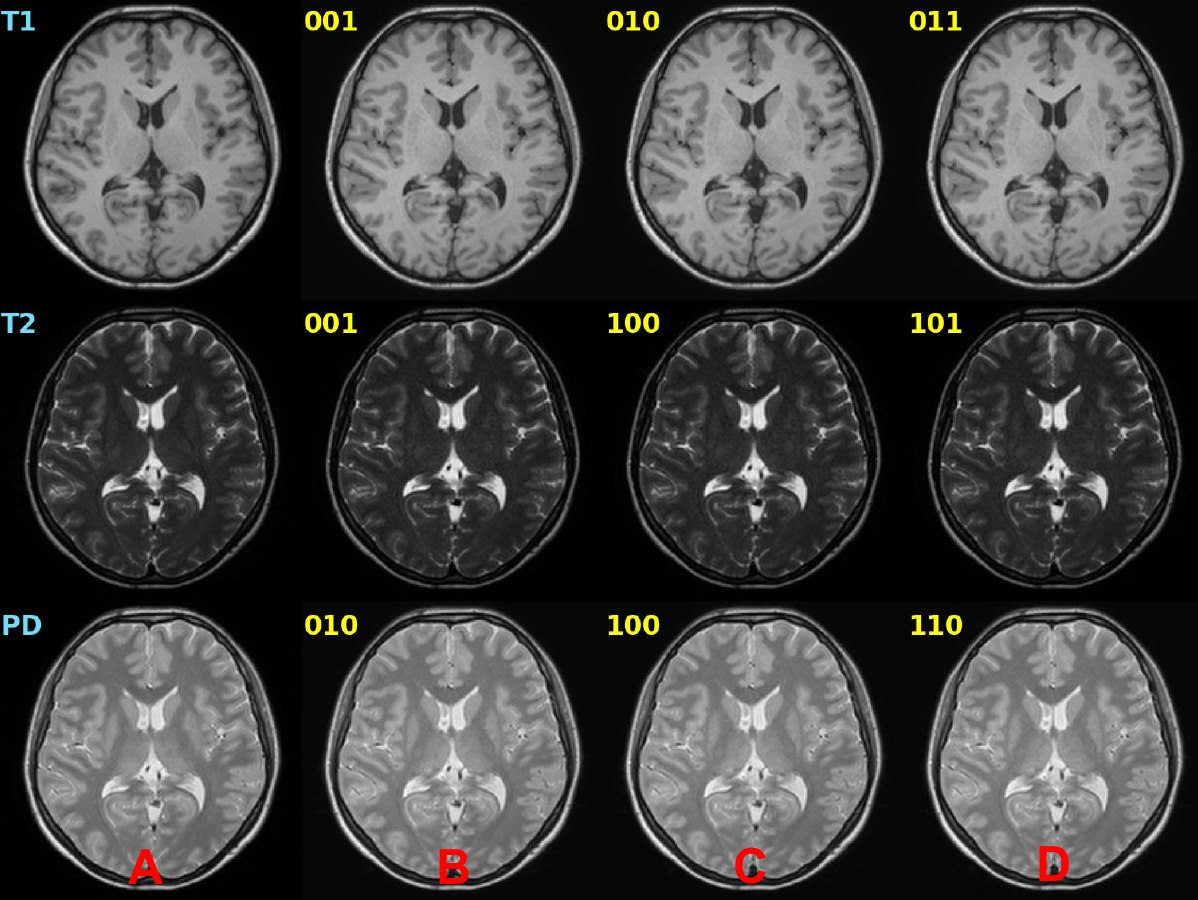}
  \caption{Qualitative results of MMT-random model on the IXI dataset. Column A: ground truth images. Column B-D: output images with input-output combination denoted by the binary string. The bits in the binary string are in the order [T1, T2, PD]. Bit value `0'/`1' means the contrast was missing/present during synthesis respectively. E.g., the binary string \textbf{101} in Row 2, Column D means the displayed T2 image was synthesized with T1 (bit 1) and PD (bit 3) as inputs.}
  \label{fig:mmt_random_bits_ixi}
\end{figure}
Table~\ref{tab:random_ixi} and~\ref{tab:random_brats} summarize the detailed performance of MMT random models for all possible input combinations, and Fig.~\ref{fig:mmt_random_bits_ixi} and~\ref{fig:mmt_random_bits_brats} show qualitative results. MMT random models can reliably synthesize the missing contrasts across different input combinations. The synthesis performance for a particular contrast improves with more input contrasts, indicating that MMT can effectively utilize the complementary information in the inputs for accurate synthesis.

\begin{figure*}[htbp]
  \centering
  \includegraphics[width=0.95\linewidth]{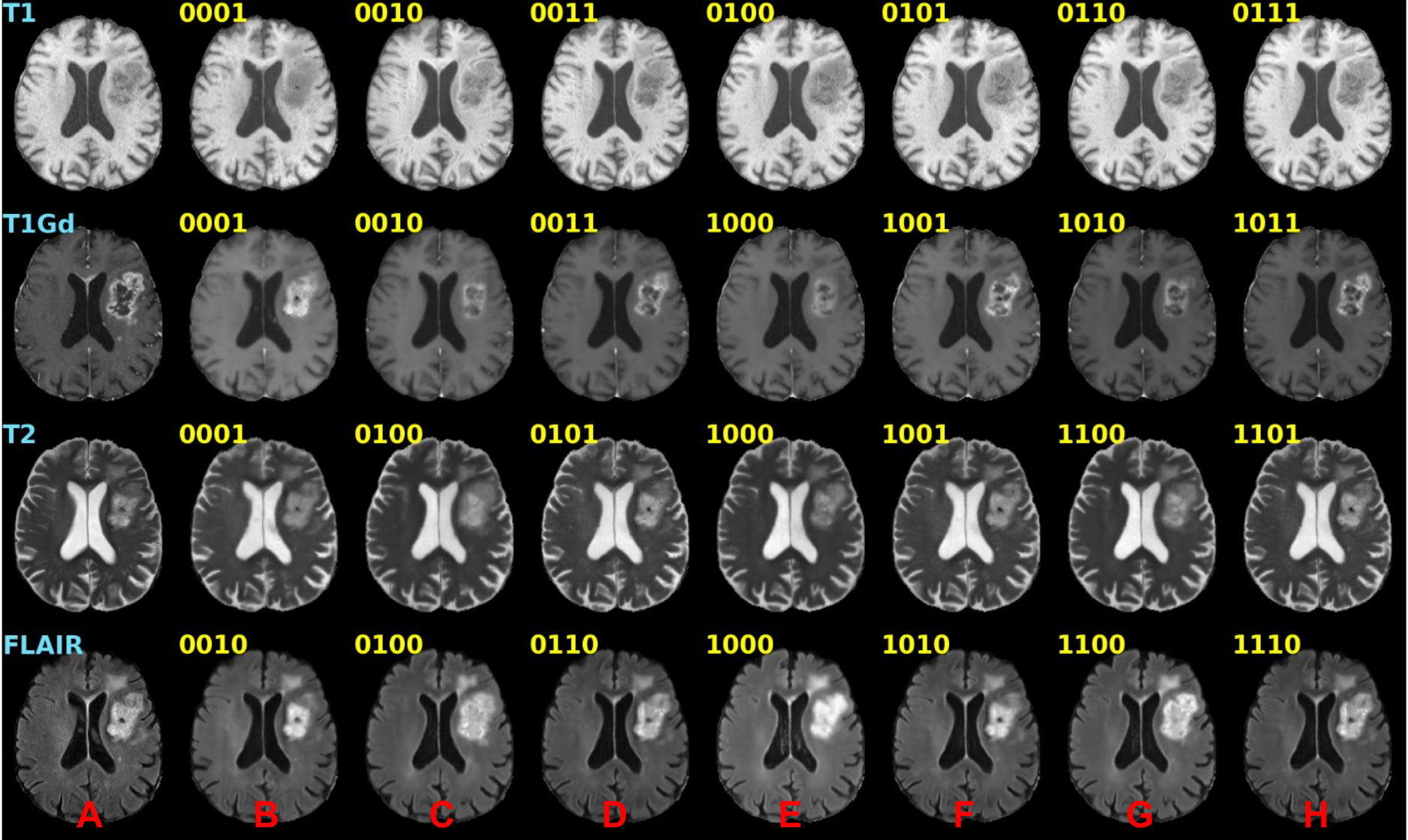}
  \caption{Qualitative results of MMT random model on the BraTS dataset. Column A: ground truth images. Column B-H: output images with input-output combination denoted by the binary string. The bits in the binary string are in the order [T1, T1Gd, T2, FLAIR]. Bit value `0'/`1' means the contrast was missing/present during synthesis respectively. E.g., the binary string \textbf{1001} in Row 2, Column F means the displayed T1Gd image was synthesized with T1 (bit 1) and FLAIR (bit 4) as inputs. The same \textbf{1001} scenario is shown for T2 synthesis in Row 3, Column F.}
  \label{fig:mmt_random_bits_brats}
\end{figure*}
\begin{figure*}[htbp]
  \centering
  \includegraphics[width=0.95\linewidth]{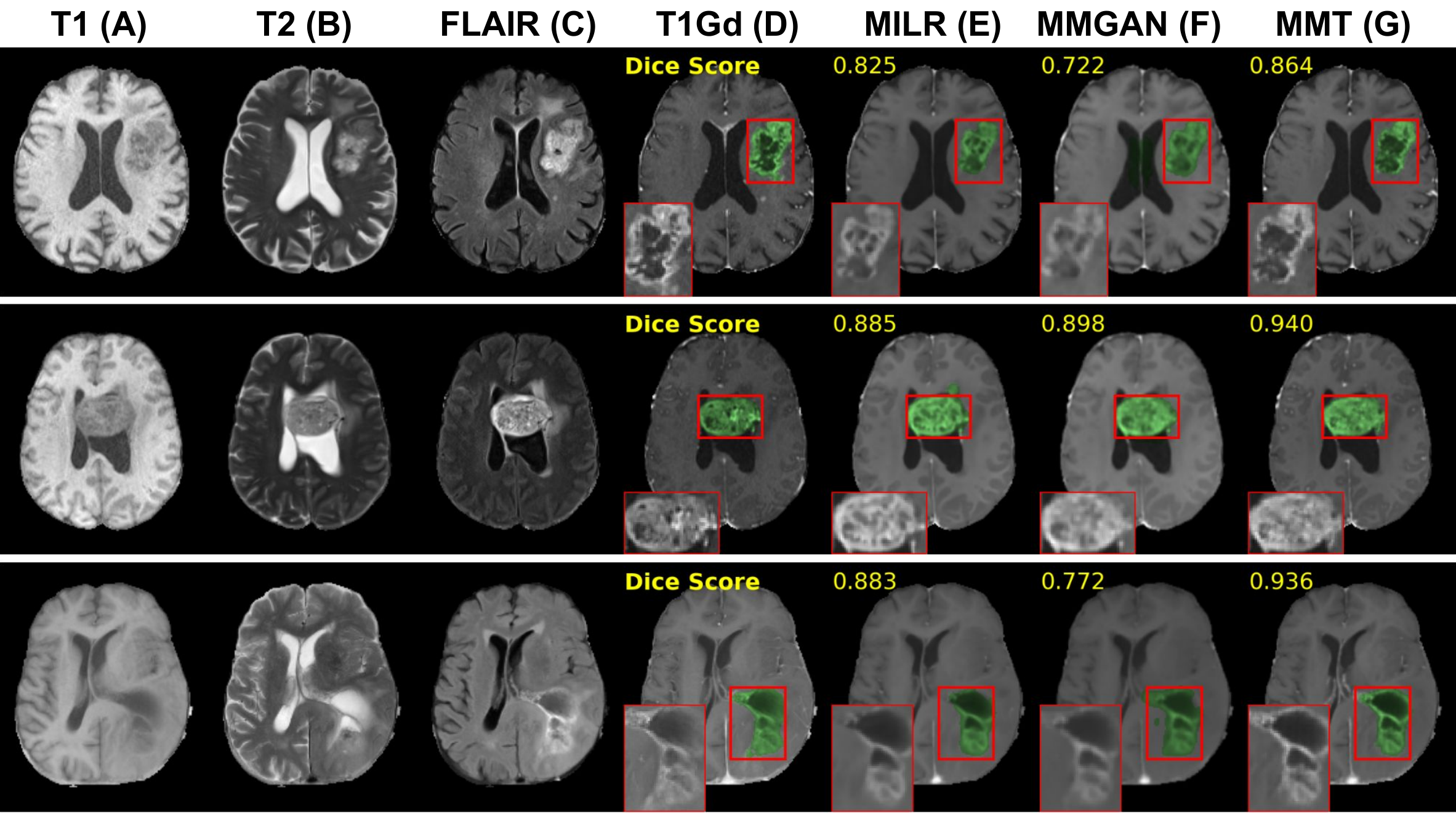}
  \caption{Comparisons on the T1Gd synthesis task on the BraTS dataset using the single models. Column A-C: input images;  Column D: ground truth T1Gd images; Column E-G: synthetic T1Gd images generated by different methods. The green overlay is the Tumor Core mask segmented on the respective images using an automatic tumor segmentation model~\cite{brats2018}. The Dice score was computed between the masks generated on ground-truth images and synthetic images.}
  \label{fig:zero_gad_comparison}
\end{figure*}

We provide a visual comparison of different single models on the T1Gd synthesis task (T1, T2, FLAIR $\rightarrow$ T1Gd) in Fig.~\ref{fig:zero_gad_comparison}. This task is of great clinical value as synthesizing post-contrast T1 images from pre-contrast images can potentially reduce the cost of post-contrast MRI, avoid adverse reactions of contrast agents, and benefit patients who may have contraindications to contrast agents, such as renal failure or allergies~\cite{mrm_gad}. As shown in Fig.~\ref{fig:zero_gad_comparison}, the outputs of MMT have better visual quality and more accurate synthesis of contrast enhancements. The enhancing tumor regions have sharper boundaries in MMT outputs compared to the other two methods. In addition, the MMT images achieve higher Dice scores when used for tumor segmentation (Sec.~\ref{sec:tumor}).

\subsection{Tumor Segmentation}
\label{sec:tumor}
\begin{figure}[htbp]
  \centering
  \includegraphics[width=0.9\linewidth]{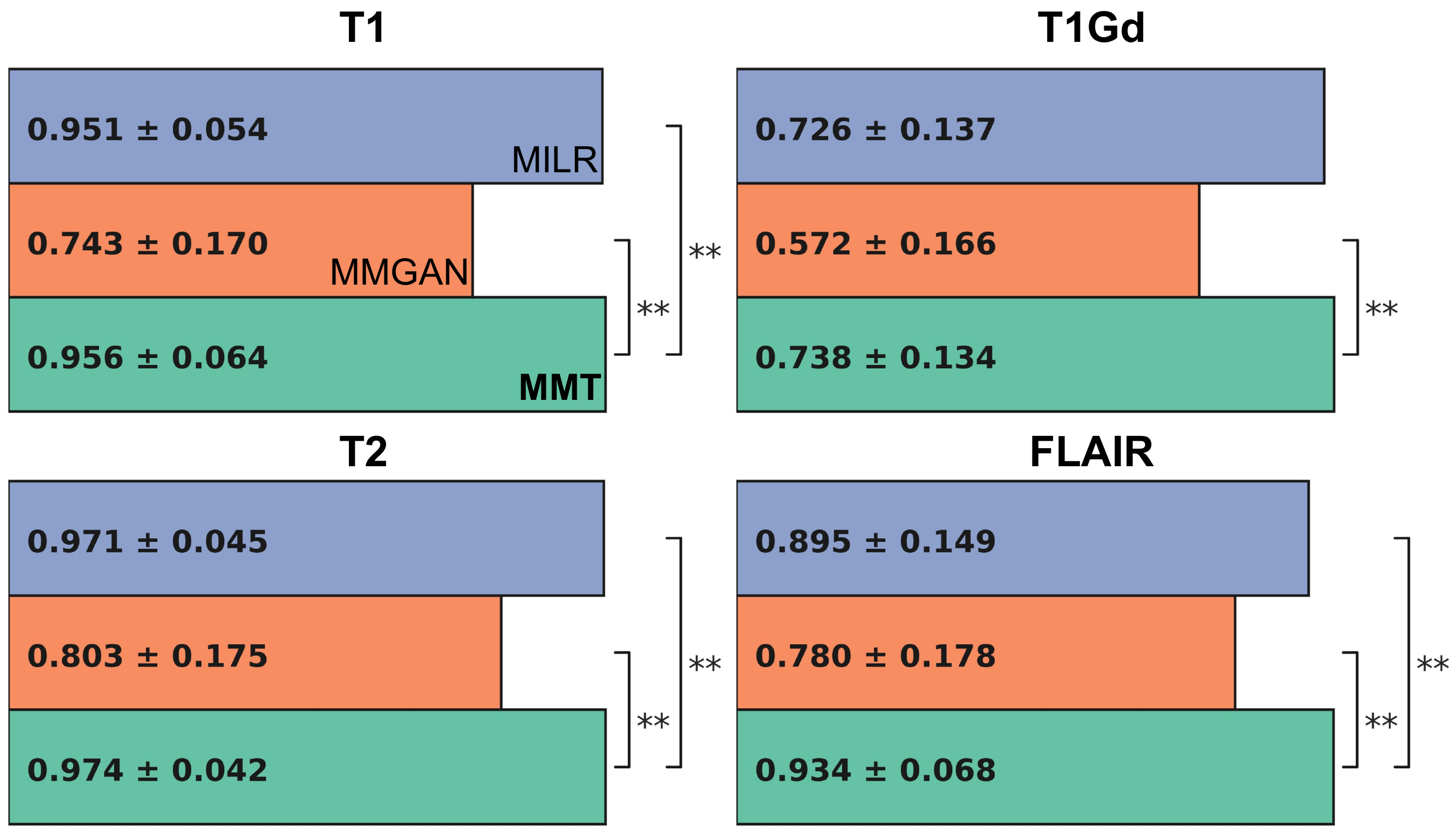}
  \caption{\looseness -1 Dice scores of different imputed sequences using a pre-trained tumor segmentation model. $\star\star$ indicates statistical significance using the Wilcoxon signed-rank test with $p < 0.005$.}
  \label{fig:brats_dice}
\end{figure}

In order to demonstrate the diagnostic equivalence between synthetic and ground truth images and for further quantitative evaluation, we perform tumor segmentation using the imputed images on the BraTS dataset. Specifically, we use the top-performing algorithm \cite{brats2021_winner} in the BraTS 2021 challenge that takes as input T1, T1Gd, T2, and FLAIR images and generates tumor masks for whole tumor (WT), enhancing tumor (ET), and tumor core (TC). We consider the single missing contrast scenario and impute the missing contrast using the single models to obtain an imputed sequence of all four required contrasts for tumor segmentation. We computed the Dice score between the segmentation mask generated from the ground truth image sequence and different imputed image sequences. We report the average Dice score of WT, ET, and TC. The results are shown in Fig.~\ref{fig:brats_dice}. MMT achieves the highest Dice scores, which are higher than 0.9 in most cases, indicating that MMT images have a more accurate depiction of the tumor regions and are highly similar to ground-truth images for diagnosis. The Dice scores of T1Gd imputed sequences are much lower than the others, suggesting that T1Gd is more challenging to synthesize than the other contrasts.

\begin{figure*}[htbp]
  \centering
  \includegraphics[width=0.82\linewidth]
  {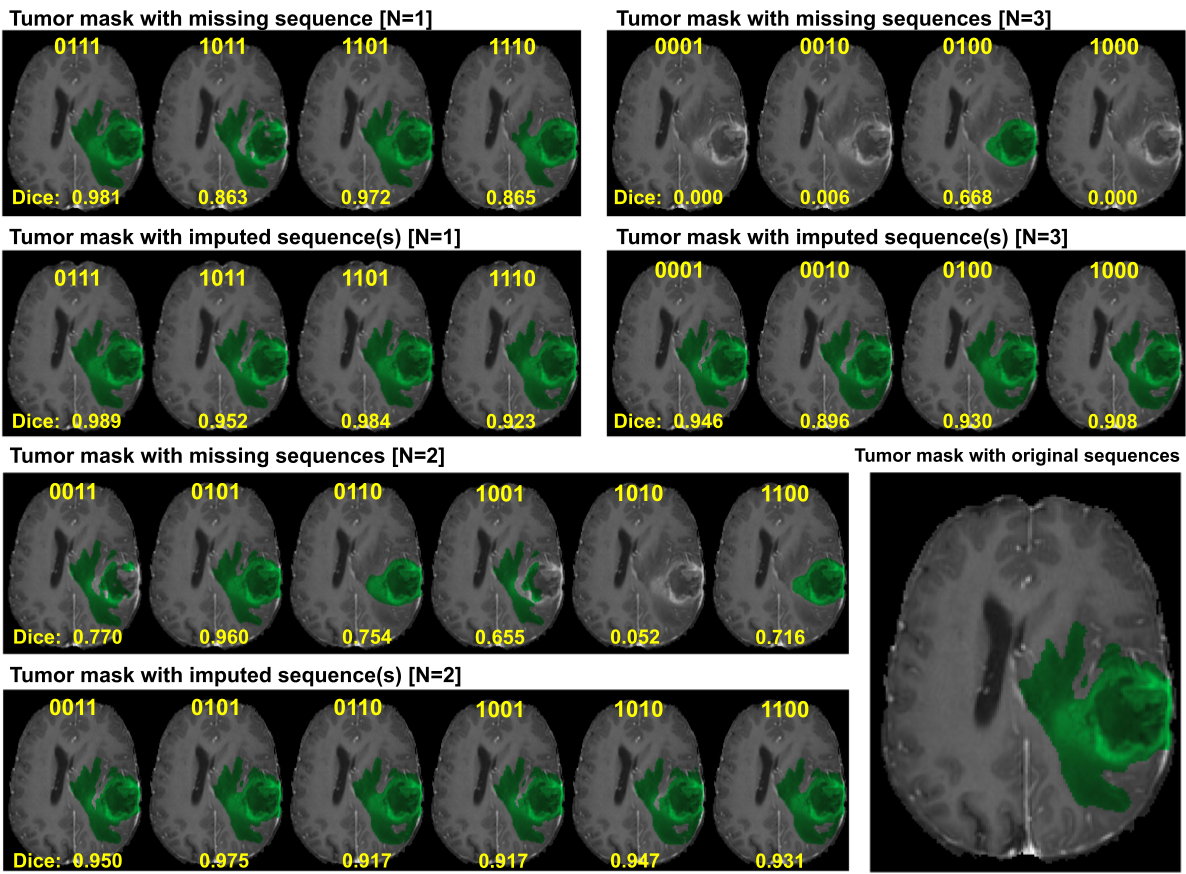}
  \caption{\looseness -1 \revise{
    Example of tumor segmentation with missing sequences and tumors segmented with sequences imputed with MMT, shown for a single case along with the corresponding Dice scores. All tumor masks are overlaid on T1Gd images for illustration purposes. The tumor mask of the original sequences is shown at the bottom right of the figure. Here, we follow the same bit-notation as Fig.~\ref{fig:mmt_random_bits_brats}.
  }}
  \label{fig:dice_ablation}
\end{figure*}

\looseness -1 \revise{
Additionally, in order to show the effectiveness of the proposed MMT model, we simulated a missing data scenario on a given tumor segmentation task as follows. Let $mask_{ori}$ be the segmented tumor using all four original input sequences. Let $mask_{miss}$ be the segmented tumor with random missing sequences. In order to generate $mask_{miss}$ for different scenarios, we trained additional tumor segmentation models that accepted only subsets of input sequences. Let $mask_{mmt}$ be the segmented tumor with the corresponding missing sequences imputed with MMT. In Table ~\ref{tab:dice_ablation} we show that 
\begin{equation}
    Dice(mask_{ori}, mask_{mmt}) > Dice(mask_{ori}, mask_{miss})
    \label{eq_dice_ablation}
\end{equation}
which implies that the images synthesized by MMT can supplement the information needed to perform clinical tasks such as tumor segmentation. Qualitative results are shown for a single case in Fig.~\ref{fig:dice_ablation}.
}

\begin{figure}[h]
  \centering
  \includegraphics[width=0.49\textwidth]{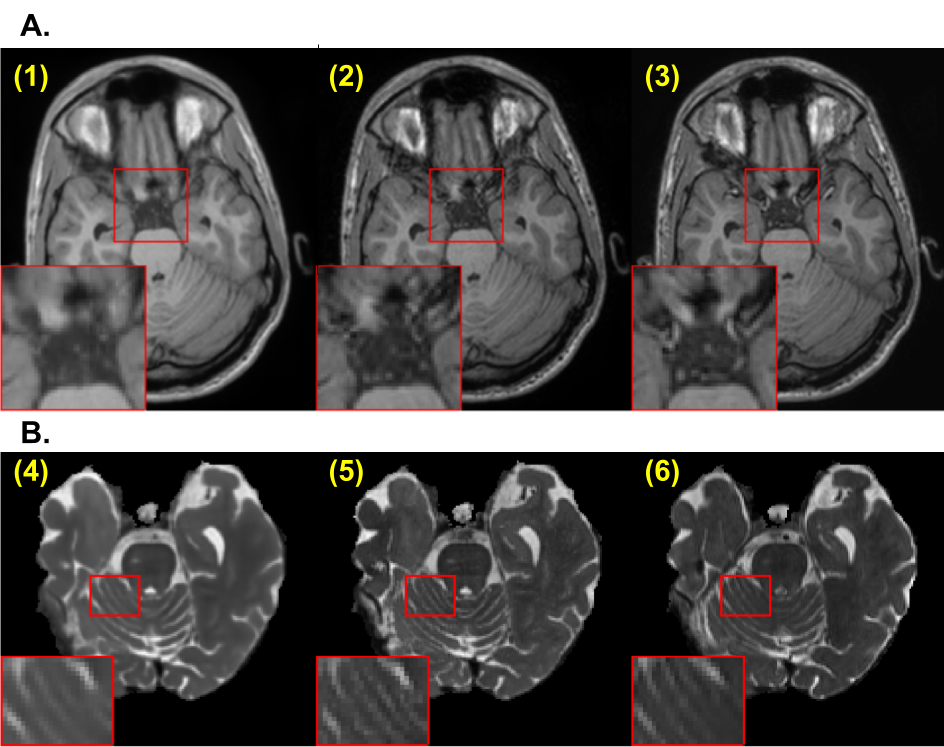}
  \caption{\revise{A: (1) output of MMT without image encoding and decoding blocks; (2) output of MMT model; (3) ground truth T1 image. Highlighted regions in the bottom left inset shows that the image quality of MMT with the CNN blocks is closer to that of the ground truth. B: (4) output of MMT without adversarial (GAN) loss; (5) output of the proposed MMT model; (6) ground truth T2 image shown for a T2 synthesis task from an MMT single model. Highlighted regions in the bottom left inset shows that the image quality of MMT (with GAN loss) is closer to that of the ground truth.}}
  \label{fig:cnn_ablation}
\end{figure}

\begin{table}
\caption{\revise{Dice score comparison between tumors segmented with missing sequences and tumors segmented with sequences imputed with MMT, relative to the tumors segmented with the original sequences. The bit value `0'/`1' in Column 2 means the contrast was missing/present for tumor segmentation, respectively. The bits in the binary string are in the order [T1, T1Gd, T2, FLAIR]. E.g., Row 6, Column 2 has the value `0101', which means that the Dice score shown in Column 3 is computed without T1 and T2 images, and the Dice score shown in Column 4 is computed with T1 and T2 imputed by the MMT model. $\star\star$ indicates statistical significance using the Wilcoxon signed-rank test with $p < 0.0005$.}}
\label{tab:dice_ablation}
\centering
\begin{tabular}{|c|c|c|c|}
\hline
\textbf{\begin{tabular}[c]{@{}l@{}}No. of Sequences \\ missing/imputed\end{tabular}} & \textbf{\begin{tabular}[c]{@{}l@{}}Missing or \\ imputed \\ sequence\end{tabular}} & \textbf{\begin{tabular}[c]{@{}l@{}}Mean Dice \\ with missing \\ sequence\end{tabular}} & \textbf{\begin{tabular}[c]{@{}l@{}}Mean Dice \\ with sequence \\ imputed \\ by MMT $\star\star$ \end{tabular}} \\ \hline
\multirow{4}{*}{1} & 0111 & $0.981 \pm 0.024$ & $0.990 \pm 0.010$ \\ \cline{2-4} 
 & 1011 & $0.957 \pm 0.049$ & $0.969 \pm 0.042$ \\ \cline{2-4} 
 & 1101 & $0.953 \pm 0.082$ & $0.976 \pm 0.052$ \\ \cline{2-4} 
 & 1110 & $0.681 \pm 0.227$ & $0.925 \pm 0.079$ \\ \hline
\multirow{6}{*}{2} & 0011 & $0.952 \pm 0.044$ & $0.967 \pm 0.042$ \\ \cline{2-4} 
 & 0101 & $0.894 \pm 0.154$ & $0.973 \pm 0.060$ \\ \cline{2-4} 
 & 0110 & $0.591 \pm 0.266$ & $0.924 \pm 0.079$ \\ \cline{2-4} 
 & 1001 & $0.844 \pm 0.191$ & $0.957 \pm 0.062$ \\ \cline{2-4} 
 & 1010 & $0.518 \pm 0.313$ & $0.914 \pm 0.087$ \\ \cline{2-4} 
 & 1100 & $0.419 \pm 0.257$ & $0.929 \pm 0.086$ \\ \hline
\multirow{4}{*}{3} & 0001 & $0.184 \pm 0.276$ & $0.956 \pm 0.070$ \\ \cline{2-4} 
 & 0010 & $0.471 \pm 0.318$ & $0.913 \pm 0.091$ \\ \cline{2-4} 
 & 0100 & $0.364 \pm 0.253$ & $0.928 \pm 0.093$ \\ \cline{2-4} 
 & 1000 & $0.084 \pm 0.177$ & $0.918 \pm 0.093$ \\ \hline
\end{tabular}
\end{table}

\revise{
\subsection{Radiomics Feature Analysis}
In order to further analyze the content of the tumor for diagnostic equivalence, we used the open-source package PyRadiomics \cite{pyrad} to compute about 120 radiomics features on the segmented tumor regions of the T1Gd ground-truth and the synthesized (zero-gad) images. These features were grouped into seven feature classes: first-order statistics, shape-based, gray level co-occurrence matrix (GLCM), gray level run length matrix (GLRLM), gray level size zone matrix (GLSZM), neighboring gray-tone difference matrix (NGTDM), and gray level dependence matrix (GLDM). The pixel values were normalized and binned using a bin width of 25, as recommended by the package. We used the mean concordance correlation coefficient (CCC) \cite{pyrad_ccc1} \cite{pyrad_ccc2} to establish the diagnostic equivalence of the segmented tumor regions. The CCC mean and standard deviation values for the seven feature classes are shown in Table ~\ref{tab:radiomics}. It can be seen that there is a high correlation between the radiomics features, which further establishes the diagnostic utility of the synthetic images. Schiewer \etal \cite{radiomic_analysis} achieved a similar range of CCC values for the assessment of the repeatability of radiomics features on small prostate tumors using test-retest Multiparametric Magnetic Resonance Imaging (mpMRI).
}

\begin{table}
\caption{\revise{Mean concordance correlation coefficient (CCC) shown for the 120 radiomic features grouped into seven feature classes. The features were computed on the segmented tumor regions on the ground truth and synthetic (zero-gad) T1Gd images. }}
\label{tab:radiomics}
\centering
\begin{tabular}{|l|c|}
\hline
\textbf{Feature Class} & \textbf{CCC} \\ \hline
First Order & $0.803 \pm 0.080$ \\ \hline
GLCM & $0.733 \pm 0.023$ \\ \hline
GLDM & $0.846 \pm 0.051$ \\ \hline
GLRLM & $0.786 \pm 0.066$ \\ \hline
GLSZM & $0.806 \pm 0.059$ \\ \hline
NGTDM & $0.843 \pm 0.112$ \\ \hline
Shape & $0.919 \pm 0.052$ \\ \hline
\end{tabular}
\end{table}

\subsection{Ablation Studies}
\label{sec:ablation}

\subsubsection{CNN Blocks}
\revise{
As described in Sec. \ref{sec:cnn}, we use shallow CNN blocks as image encoding and decoding blocks in MMT. We validate the effectiveness of this design by removing these blocks from MMT. We train the ablated model on the IXI dataset in the \textit{single missing contrast} scenario, and Table ~\ref{tab:ablation_cnn} summarizes the quantitative results. Removing the CNN blocks results in similar SSIM and slightly lower PSNR scores in general but much higher LPIPS scores, which indicates much lower perceptual quality. Fig.~\ref{fig:cnn_ablation} shows a comparison of output images. The output of ablated models has a blocky appearance since the Transformer blocks are based on local windows. By adding the image encoding and decoding blocks, the output is more continuous and has better visual quality.
}

\begin{table}
\caption{\revise{Quantitative metrics on the IXI dataset in the \textit{single missing contrast} scenario. ``w/o CNN" is the ablated MMT model without CNN encoding and decoding blocks}}
\label{tab:ablation_cnn}
\centering
\begin{tabular}{l|l|ccc} 

\hline
Metric                 & Method & T1             & T2             & PD              \\ 
\hline
\multirow{2}{*}{SSIM $\uparrow$}  & MMT   & 0.936 & 0.976 & \textbf{0.978} \\
                      & ~~w/o CNN    & \textbf{0.939} & \textbf{0.977} & 0.976 \\ 
\hline
\multirow{2}{*}{PSNR (dB) $\uparrow$}  & MMT  & 34.60 & \textbf{34.31} & \textbf{40.85}   \\
                      & ~~w/o CNN    &  \textbf{35.00} & 34.17 & 40.49 \\ 
\hline
\multirow{2}{*}{LPIPS $\downarrow$} & MMT   & \textbf{0.096}  & \textbf{0.072}  & \textbf{0.069}   \\
                      & ~~w/o CNN    &  0.120 & 0.119 & 0.086\\
\hline
\end{tabular}
\end{table}

\subsubsection{Cross-contrast Attention}
\revise{
In order to understand the effect of cross-contrast attention computation in the M-Swin block, we trained the MMT Single model on the BraTS dataset without cross-contrast attention. In this training mode, the contrast dimension in the window attention was collapsed to treat all the contrast tokens in the same manner effectively. Although the quantitative performance of this model was almost similar to that of MMT, the lack of cross-contrast attention computation affected the model's interpretability. It can be seen in Fig. \ref{fig:attn}-B that the attention heat maps are randomly activated on the different contrasts, whereas the heat maps of the model with cross-contrast attention are active on the areas of lesions on the respective contrast images. This shows that cross-contrast attention computation is crucial for the explainability and interpretability of the proposed model.
}

\subsubsection{Adversarial Loss}
\revise{To show the effect of adversarial loss on the model performance, we trained the MMT Single model on the BraTS dataset without the adversarial (GAN) loss. Fig. \ref{fig:cnn_ablation}-B shows that the model with GAN loss (image 5) has well-defined branches in the cerebellum region when compared to the model without GAN loss (image 4). Table~\ref{tab:ablation_gan} shows the average quantitative metrics for the single missing contrast scenario for this ablation study. Although the pixel-level metrics like SSIM and PSNR are marginally better for the model without GAN loss, the LPIPS perceptual metric is consistently better for the proposed MMT model.}

\begin{table}
\caption{\revise{Quantitative metrics on the BraTS dataset in the \textit{single missing contrast} scenario. ``w/o GAN" is the ablated MMT model without adversarial (GAN) loss.}}
\label{tab:ablation_gan}
\centering
\begin{tabular}{l|l|cccc} 

\hline
Metric                 & Method & T1             & T1Gd             & T2 & FLAIR              \\ 
\hline
\multirow{2}{*}{SSIM $\uparrow$}  & MMT   & 0.941 & \textbf{0.939} & 0.939 & 0.911 \\
                      & ~~w/o GAN    & \textbf{0.947} & 0.938 & \textbf{0.941} & \textbf{0.912} \\ 
\hline
\multirow{2}{*}{PSNR (dB) $\uparrow$}  & MMT  & 28.06 & 29.74 & 27.13 & \textbf{26.30}  \\
                      & ~~w/o GAN    &  \textbf{28.29} & \textbf{29.82} & \textbf{27.30} & 26.24 \\ 
\hline
\multirow{2}{*}{LPIPS $\downarrow$} & MMT   & \textbf{0.081}  & \textbf{0.120}  & \textbf{0.086} & \textbf{0.125}   \\
                      & ~~w/o GAN    &  0.105 & 0.127 & 0.093 & 0.130\\
\hline
\end{tabular}
\end{table}

\subsection{Attention Score Analysis}
\label{sec:att}
\begin{figure}[htbp]
  \centering
  \includegraphics[width=1.0\linewidth]{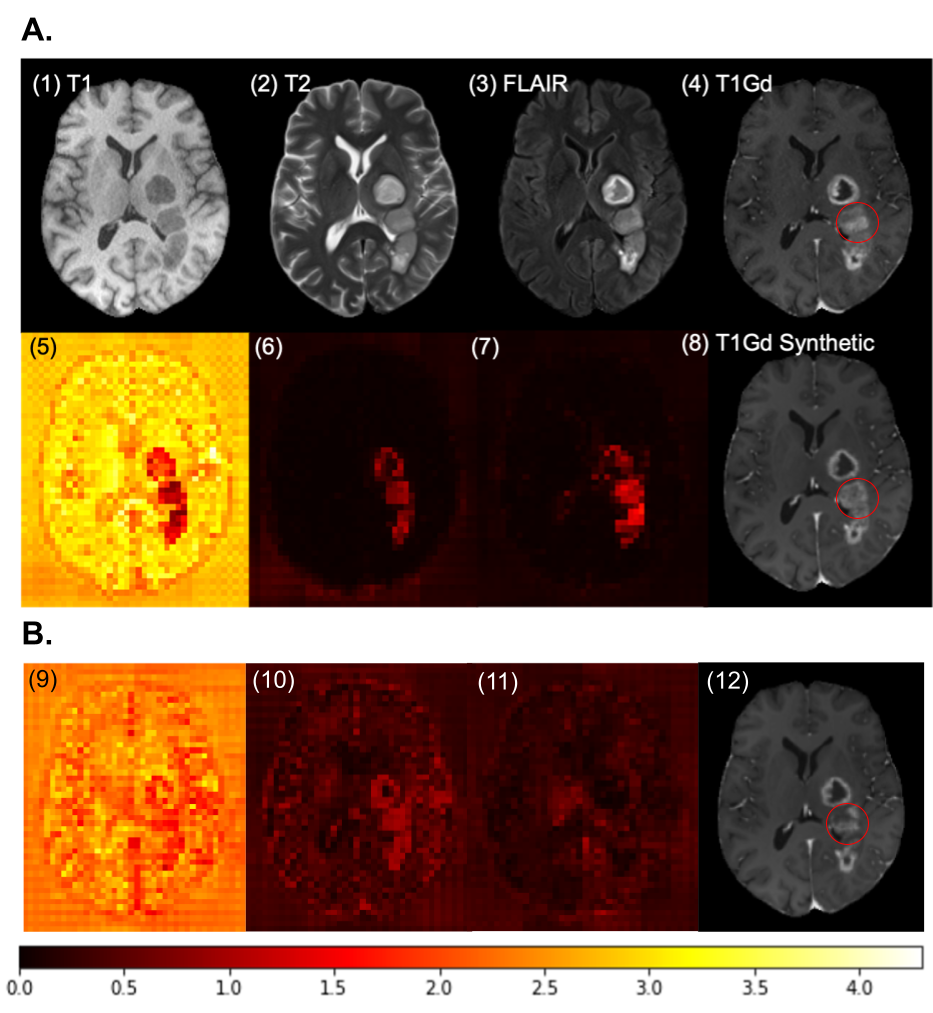}
  \caption{(A) Attention score visualization. (1)-(3): input images; (4) ground-truth T1Gd image; (5)-(7): attention scores corresponding to inputs (1)-(3) from the last M-Swin decoder block; (8) output T1Gd image. \revise{(B) Attention score visualization from the model trained without cross-contrast attention in the M-Swin blocks of encoder and decoder. The attention heat map from (A) is more explainable when compared to that from the model without cross-contrast attention. The synthesized T1Gd from the model with cross-attention (8) is better than the T1Gd synthesized without cross-attention (12), especially in the area highlighted by the red circle.}}
  \label{fig:attn}
\end{figure}

The attention scores inside the MMT decoder indicate the amount of information coming from different input contrasts and regions for synthesizing the output, which makes MMT inherently interpretable.
We show an example of attention score visualization in Fig.~\ref{fig:attn}-A. A higher attention score indicates more information coming from a particular region. To synthesize the T1Gd image, MMT mostly looks at the T1 input, which is
reasonable since T1 and T1Gd are very similar except for the enhancing regions in T1Gd. However, for the tumor region, MMT extracts more information from T2 and FLAIR as they provide stronger signals for the lesion. This demonstrates that MMT understands the image context and the input-target contrast relationship and pays attention to the appropriate regions and contrasts for synthesis. 

\begin{figure}[htbp]
    \centering
    \includegraphics[width=0.8\linewidth]{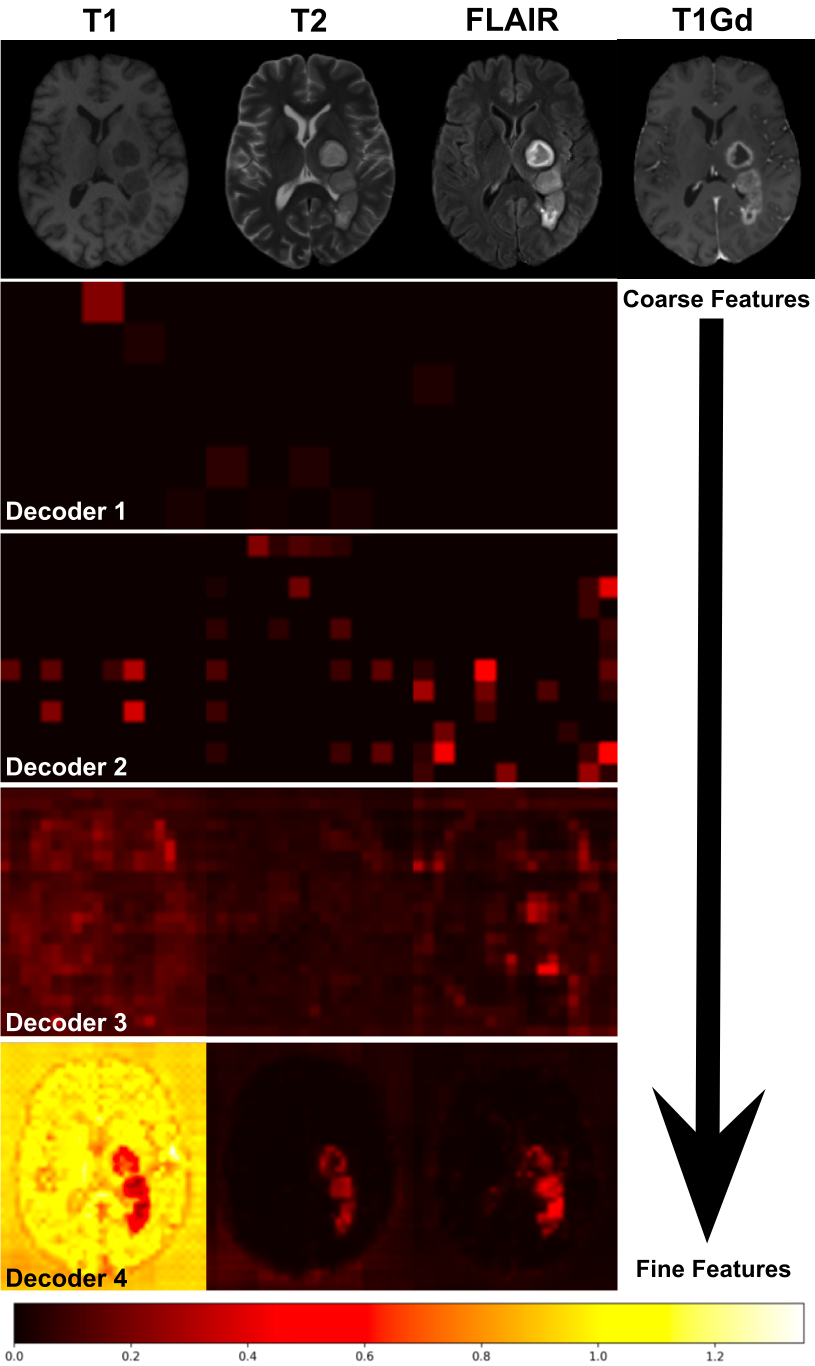}
    \caption{\revise{Attention visualization for individual decoder layers across individual input contrasts. It can be seen that the features learned by the decoder are in a coarse-to-fine fashion. The example shown here is the same as Fig.~\ref{fig:attn}. }}
    \label{fig:dec_attn}
    \vspace{-2mm}
\end{figure}

\revise{
Additionally, to show that the MMT decoder progressively decodes the encoder outputs at different scales and generates the output image in a coarse-to-fine fashion, we visualized the attention from all four decoder layers and for all the input contrasts for the same example as above. It can be seen from Fig.~\ref{fig:dec_attn} that the features in the initial decoder layers are coarse, and the features progressively become finer and focus on the area of lesion to synthesize the T1Gd image. 
}

\begin{figure}[htbp]
  \centering
  \includegraphics[width=0.95\linewidth]{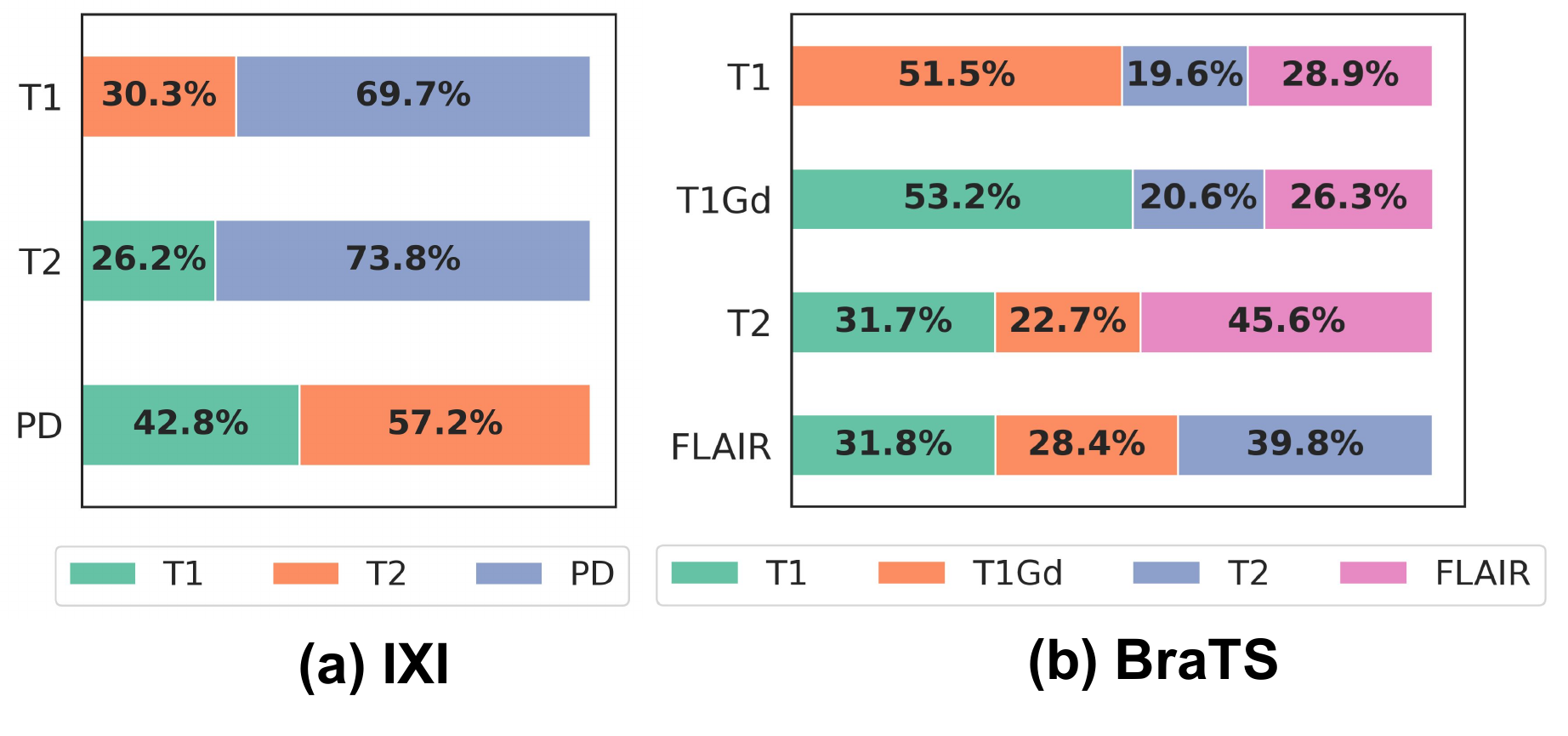}
  \caption{Percentage of attention scores of input contrasts.}
  \label{fig:attn_pct}
\end{figure}

We can \revise{also} quantitatively measure the relative importance of each input contrast for a particular output by the percentage of attention scores. Specifically, for each input contrast, we sum the attention scores overall MMT decoder blocks and compute the percentage of attention scores each input conveys. These percentages quantify the amount of information contributed by each input image and indicate their relative importance. We utilize the MMT single models and average the scores on the test sets. The results are shown in Fig.~\ref{fig:attn_pct}. On the IXI dataset, PD is the most important input for synthesizing T1 or T2 since it contributes most of the information ($\sim$70\%). For synthesizing PD, T2 contributes slightly more information than T1, which suggests a higher similarity between T2 and PD. For the BraTS dataset, T1 and T1Gd are essential inputs for each other, contributing $\sim$50\% of the information, which is reasonable since T1 and T1Gd are very similar except for the contrast-enhancing regions in T1Gd. Similarly, T2 and FLAIR are the most critical input for each other, contributing $\sim$40\% of the information, which is consistent given that they are quite similar, except for the suppression of cerebrospinal fluid for FLAIR.

\section{Discussions and Conclusion}
\looseness -1    In this work, we propose a multi-contrast multi-scale Transformer for the synthesis of missing MRI contrasts. MMT overcomes the limitations of CNN models, such as the requirement for a fixed number of input and output channels, the inability to capture long-range dependencies, and the lack of interpretability. Our extensive experiments on the IXI and BraTS datasets demonstrate that MMT achieves better quantitative and qualitative performance than the state-of-the-art methods. We further demonstrate the diagnostic equivalence of MMT-imputed and ground-truth images on the tumor segmentation task. Attention score analysis reveals interesting and reasonable insights into the relative importance of different input contrasts and regions to synthesize a particular image contrast. \revise{Even though in some cases, the pixel-based metrics like SSIM/PSNR are not significantly different between MMT and other competing methods (Tables \ref{tab:random_ixi}, \ref{tab:random_brats}, and \ref{tab:ablation_gan}), the LPIPS perceptual quality metric is consistently superior for the proposed MMT model. This shows that the MMT model has better overall perceptual quality despite having similar pixel-based quality metrics.}
    
    The proposed work has several potential clinical applications. MMT not only serves as a means of synthesizing missing contrasts but also has the potential to make the overall acquisition process more efficient and safer. \revise{This method could be used to avoid Gadolinium administration in patients with contraindications such as a susceptibility to nephrogenic systemic fibrosis or in pediatric patients. Synthesizing accurate T1Gd images from pre-contrast images is an extremely challenging task~\cite{postgad1, postgad2} and there are no state-of-the-art methods that perform well. Hence, the results shown in this work are a significant step toward accurate zero-gad synthesis.} The model could also synthesize redundant contrasts, such as STIR in lumbar spine exams, using other contrasts, such as T1 and T2. The proposed algorithm can also avoid quality and motion issues during the scan by imputing the corrupted image from other good-quality sequences and avoiding needing to call patients for extra scanning to acquire missing or corrupted sequences.
    
    Future work could feasibly extend our proposed methods to cross-modality synthesis or image segmentation. For example, it can be used for cross-modality synthesis such as MRI $\rightarrow$ CT. Beyond image synthesis, it can also be applied to image segmentation by adding a ``segmentation query" in the MMT decoder. This way, we can train a segmentation model operating on missing input contrasts. The image segmentation and synthesis tasks can also be trained jointly using multi-task learning. 

    \begin{figure}[t]
      \centering
      \includegraphics[width=1.0\linewidth]{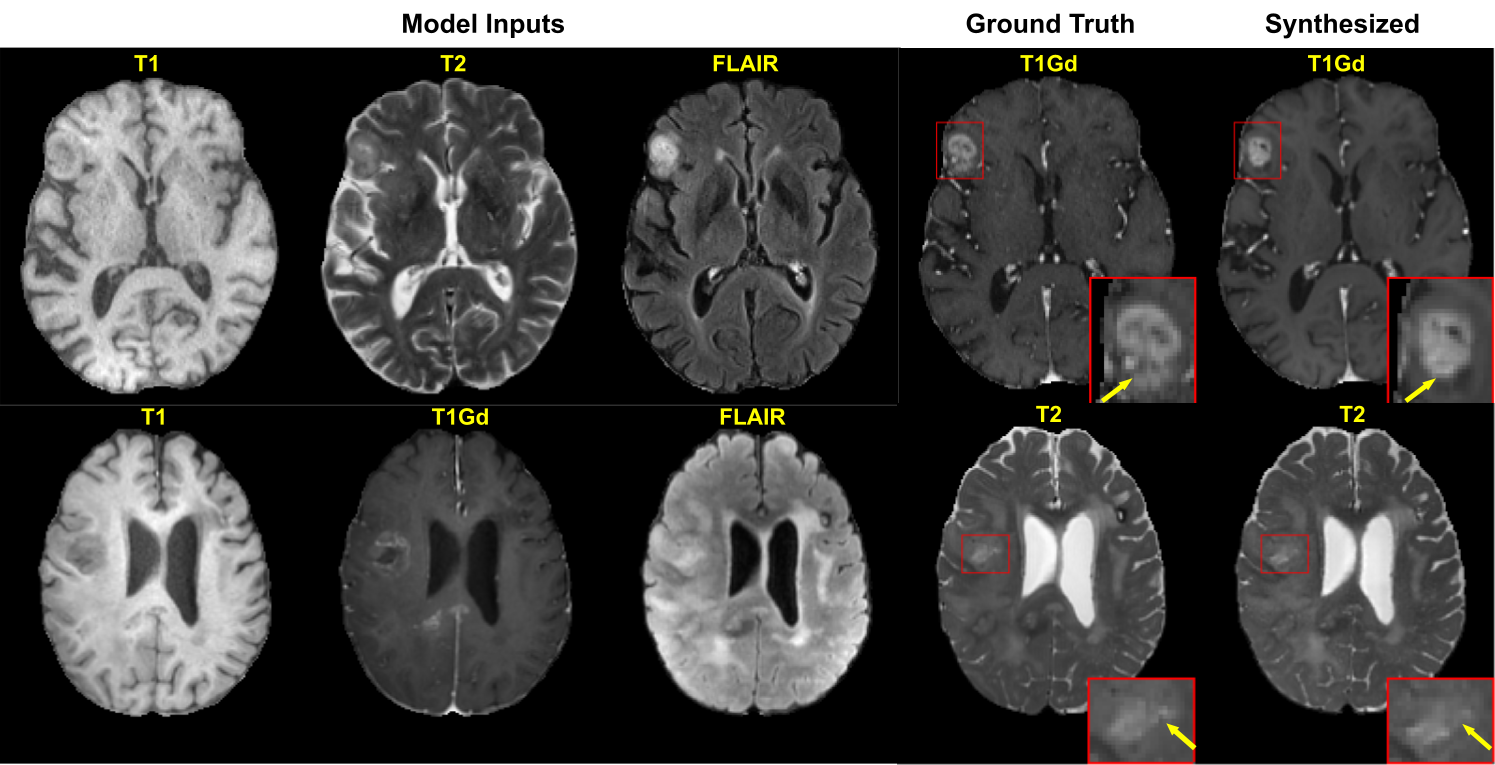}
      \caption{\revise{Examples of two cases where the enhancement patterns in the synthesized images are not the same as that of the ground truth as pointed out by yellow arrows. The top and bottom rows are examples of T1Gd synthesis and T2 synthesis, respectively, using the remaining three sequences as inputs.}}
      \label{fig:failure_mode}
    \end{figure}

    One of the major limitations of this work is the need for image registration for multi-contrast inputs, which can be time-consuming depending on the anatomy. \revise{This also means that all input contrasts must be in the same orientation (AX/SAG/COR). The lack of a reliable registration method will lead to poor performance of the proposed model. In a few cases, we noticed that the enhancement or lesion patterns in the synthesized images were not an exact match with that of the ground truth as shown in Fig.~\ref{fig:failure_mode}. To assess such differences and their effect on the diagnostic outcome,} tumor segmentation is a reasonable proxy but an extensive reader study is needed for subjective evaluation and clinical performance, which is beyond the scope of this work. A multi-center evaluation is needed to assess the model's generalizability across different sites, scanners, and clinical settings. \revise{The advantages of Transformer based networks come with the overhead of memory footprint and increased number of parameters. The model training and deployment require a significant amount of RAM and GPU memory. Transformers also require a large number of training samples, which limits the extension of this work to smaller datasets.}.

\bibliographystyle{ieeetr}
\bibliography{refs}
\end{document}